\pdfoutput=1
\documentclass[fleqn,usenatbib,useAMS]{mn2e}

\usepackage{graphicx}	

\usepackage{amsmath}	
\usepackage{amssymb}	
\usepackage{multicol}        
\usepackage{bm}		
\usepackage{pdflscape}	
\usepackage{scrextend} 
\usepackage{multirow}
\usepackage[greek,english]{babel}

\long\def\symbolfootnote[#1]#2{\begingroup%
\def\thefootnote{\fnsymbol{footnote}}\footnote[#1]{#2}\endgroup}

\newcommand{\kms}{\,km\,s$^{-1}$} 
\newcommand{\ergs}{\,erg\,s$^{-1}$} 

\newcommand{\ha}{H$\textrm{\greektext a}$~}
\newcommand{\hb}{H$\textrm{\greektext b}$~}
\newcommand{\hans}{H$\textrm{\greektext a}$}
\newcommand{\hbns}{H$\textrm{\greektext b}$}

\newcommand{\mgii}{Mg\,{\sc ii}~}
\newcommand{\mgiins}{Mg\,{\sc ii}}
\newcommand{\feii}{Fe\,{\sc ii}~}

\newcommand{\civ}{C\,{\sc iv}~}
\newcommand{\civns}{C\,{\sc iv}}

\newcommand{\ciins}{C\,{\sc ii}}

\newcommand{\heins}{He\,{\sc i}}

\newcommand{\oiv}{O\,{\sc iv}~}
\newcommand{\oivns}{O\,{\sc iv}}

\newcommand{\siivns}{Si\,{\sc iv}}
\newcommand{\oiii}{[O\,{\sc iii}]~}
\newcommand{\oiiins}{[O\,{\sc iii}]}

\newcommand{\niins}{[N\,{\sc ii}]}

\newcommand{\siins}{[S\,{\sc ii}]}

\def \ll {$\lambda\lambda$}

\usepackage[T1]{fontenc}
\usepackage{ae,aecompl}

\usepackage{txfonts}

\volume{465}

\title[BHM Correction]{Correcting \civns-Based Virial Black Hole Masses}

\author[L. Coatman et al.]{
Liam Coatman,$^{1}$\thanks{E-mail: lcoatman@ast.cam.ac.uk}
Paul C. Hewett,$^{1}$
Manda Banerji,$^{1,2}$ 
Gordon T. Richards,$^{3}$\newauthor
Joseph F. Hennawi,$^{4}$ and
J. Xavier Prochaska$^{5}$ 
\\
$^{1}$Institute of Astronomy, University of Cambridge, Madingley Road, Cambridge, CB3 0HA, UK\\
$^{2}$Kavli Institute for Cosmology, University of Cambridge, Madingley Road, Cambridge, CB3 0HA, UK\\
$^{3}$Department of Physics, Drexel University, 3141 Chestnut Street, Philadelphia, PA 19104, USA\\
$^{4}$Max-Planck-Institut f{\"u}r Astronomie, K{\"o}nigstuhl 17, D-69117 Heidelberg, Germany\\
$^{5}$UCO/Lick, University of California Santa Cruz, Santa Cruz, California, CA 95064, USA
}

\date{Accepted XXX. Received YYY; in original form ZZZ}

\pubyear{2017}

\begin{document}
\label{firstpage}
\pagerange{\pageref{firstpage}--\pageref{lastpage}}
\maketitle

\begin{abstract}
The \civns$\lambda\lambda$1548,1550 broad emission line is visible in optical spectra to redshifts exceeding $z\sim5$. 
\civ has long been known to exhibit significant displacements to the blue and these `blueshifts' almost certainly signal the presence of strong outflows.
As a consequence, single-epoch virial black hole (BH) mass estimates derived from \civ velocity-widths are known to be systematically biased compared to masses from the hydrogen Balmer lines. 
Using a large sample of 230 high-luminosity ($L_{\rm Bol} = 10^{45.5}-10^{48}$ erg s$^{-1}$), redshift $1.5 < z < 4.0$ quasars with both \civ and Balmer line spectra, we have quantified the bias in \civ BH masses as a function of the \civ blueshift. 
\civ BH masses are shown to be a factor of five larger than the corresponding Balmer-line masses at \civ blueshifts of 3000\kms and are over-estimated by almost an order of magnitude at the most extreme blueshifts, $\gtrsim 5000$\kms.
Using the monotonically increasing relationship between the \civ blueshift and the mass ratio BH(\civns)/BH(\hans) we derive an empirical correction to all \civ BH-masses.
The scatter between the corrected \civ masses and the Balmer masses is 0.24 dex at low \civ blueshifts ($\sim$0\kms) and just 0.10 dex at high blueshifts ($\sim$3000\kms), compared to 0.40 dex before the correction. 
The correction depends only on the \civ line properties - i.e. full-width at half maximum and blueshift - and can therefore be applied to all quasars where \civ emission line properties have been measured, enabling the derivation of un-biased virial BH mass estimates for the majority of high-luminosity, high-redshift, spectroscopically confirmed quasars in the literature. 
\end{abstract}

\begin{keywords}
quasars: supermassive black holes -- galaxies: evolution
\end{keywords}



\section{Introduction}

The goal of better understanding the origin of the correlation between the masses of super-massive black holes (BHs) and the masses of host-galaxy spheroids has led to much work focussing on the properties of quasars and active galactic nuclei (AGN) at relatively high redshifts, $z\gtrsim 2$. 
Extensive reverberation-mapping campaigns \citep[e.g.][]{kaspi00,kaspi07,peterson04,bentz09,denney10} have been used to calibrate single-epoch virial-mass estimates which use the velocity widths of the hydrogen Balmer emission lines and the nuclear continuum luminosity to provide reliable BH masses \citep[e.g.][]{greene05,vestergaard06,vestergaard09,shen11,shen12,trakhtenbrot12}. 
Single-epoch virial BH mass estimates using \hb are possible up to redshifts $z\sim0.7$, and the technique has been extended to redshifts $z\sim1.9$ via the calibration of the broad \mgiins$\lambda\lambda$2796,2803 emission line \citep{mclure02,onken08,wang09,rafiee11}. 
At redshifts $z\gtrsim2$, however, ground-based statistical studies of the quasar population generally have no access to the rest-frame optical and near-ultraviolet spectral regions.

Attention has thus been drawn to the properties of the \civns$\lambda\lambda$1548, 1550 emission line, which is both relatively strong in the majority of quasars and visible in modern `optical' spectra, such as those provided by the Sloan Digital Sky Surveys, to redshifts exceeding $z\sim5$. 
In contrast to a number of low-ionisation emission lines, such as \mgiins, the \civ emission has long been known to exhibit significant displacements to the blue \citep{gaskell82} and more recent work \citep[e.g.][]{sulentic00a, richards11} has established that the extent of `blueshifts' in the \civ emission correlates with a number of properties of quasar spectral energy distributions (SEDs). 
While the physical origin of the blueshifted emission has not been established there is a consensus that the associated gas is not tracing virial-induced velocities, that should allow a BH-mass estimate to be derived.  
A favoured interpretation associates the blueshifted emission with out-flowing material \citep[see][for a recent review]{netzer15}, reaching velocities significantly larger than virial-induced velocities associated with the BH \citep[e.g.][]{sulentic07, richards11}.
Certainly, excess emission-line flux in the blue wing of the \civ emission increases commonly employed measures of the line-width, notably the full-width at half maximum (FWHM) and the line dispersion ($\sigma$). 
As a consequence, BH-masses derived from \civ emission line velocity-widths are known to be systematically biased compared to masses from the Balmer lines \citep[e.g.][]{shen08,shen12,coatman16}. 

In recent literature, attempts have been made to minimise the influence of the systematic non-virial contribution to the \civ emission on estimates of the BH mass. 
Strategies include (i) significantly reducing the dependence of the derived masses on the emission-line velocity width (e.g. from the $V^2$ dependence predicted assuming a virialized broad line region to just $V^{0.56}$ in \citealt{park13}; see also \citealt{shen12}), (ii) adopting a measure of emission-line velocity-width that is relatively insensitive to changes in the core of the emission-line profile \citep[e.g.][]{denney13} and (iii) estimating the amplitude of the non-virial contribution to the \civ emission-line via comparison with other ultraviolet emission lines (e.g. \siivns+\oivns$\lambda$1400 in \citealt{runnoe13} and \citealt{brotherton15}).
The increased number of high-quality spectra of quasars where information on both the Balmer lines in the rest-frame optical and \civ in the ultraviolet is available enables a rather different approach. 
Specifically, to investigate whether, using the properties of the \civ emission line itself, it is possible to reduce, or even remove, the systematic bias in the BH-mass estimates. 

In this paper we analyse the spectra of 230 high-luminosity ($10^{45.5}-10^{48}$\,\ergs), redshift $1.5 < z < 4.0$ quasars for which spectra of the hydrogen Balmer emission lines and the \civ emission line exist. 
A direct comparison of the emission-line velocity widths is therefore possible, allowing us to determine a highly effective empirical correction to the \civ emission line velocity width as a function of the \civ emission line blueshift. 

The paper is structured as follows. Section 2 presents the extensive set of near-infrared spectra that, combined with optical spectra of the quasar ultraviolet rest-frame, provides our spectroscopic
catalogue. 
The scheme adopted to calculate emission-line parameters, which draws heavily on the methodology of \citet{shen11}, \citet{shen12} and \citet{shen16a}, is described in Section 3. 
The observational results, where the emission-line properties of the Balmer lines and the \civ emission are compared and a quantitative relationship derived, are included as
Section 4. 
Then, in Section 5, the practical application of the new BH-mass estimation formula and the extent of remaining uncertainties are discussed, and our scheme is compared to others presented in the literature. 
Finally, we summarise the main points of the paper and highlight forthcoming improvements to systemic redshift estimates for quasars that should improve the accuracy of BH-masses from rest-frame ultraviolet quasar spectra even further.
Throughout this paper we adopt a $\Lambda$CDM cosmology with $h_0=0.71$, $\Omega_M=0.27$, and $\Omega_\Lambda=0.73$. 
Vacuum wavelengths are used for both rest-frame ultraviolet and optical features.
Unless otherwise stated, optical (i.e. SDSS) magnitudes are given in the AB system and infra-red magnitudes in the Vega system, following the conventions of the original surveys. 

\section{Quasar Sample}

The aim of this work is to measure empirically the systematic bias in \civns-based virial BH mass estimates for high-$z$ quasars as a function of the \civ emission-line blueshift. 
The basis for the \civ blueshift based correction is a large sample of quasars where it is possible to make a direct comparison of the \civ line-width with the line-width of the low-ionisation Balmer lines \ha and \hbns, which are believed to provide reliable proxies for the virial velocity. 
Such an approach has not been possible hitherto as spectra that cover both the observed-frame optical (where the redshifted \civ appears) and near-infrared (where \hb and \ha lie) are required.

We have compiled a sample of 307 quasars at redshifts $1.5 < z < 4$ with both optical and near-infrared spectra to enable such a comparison to be performed. 
Reliable emission line properties were measured for 230 quasars (Section~\ref{sec:flagged_spectra}), with 164 possessing \ha line measurements and 144 \hb line measurements.  
The sample is considerably larger than previous studies of the rest-frame optical spectra of high-$z$ quasars \citep[e.g.][]{shen12}. 
As we demonstrate in Section~\ref{sec:effectiveness}, the quasars have \civ blueshifts of up to $\sim$5000\kms, and span the full range observed in the population. 
Part of this data set has been taken from the literature, but a substantial fraction is presented here for the first time. 
The infrared spectra were acquired using several different telescope and spectrograph combinations and the contributions from each telescope/spectrograph, along with the instrumental configurations, are summarised in Table~\ref{tab:specnums}. 
We have sub-divided our sample into two overlapping groups: quasars with reliable \ha line measurements (the `\ha sample') and quasars with reliable \hb measurements (the `\hb sample').

\begin{figure}
    \includegraphics[width=\columnwidth]{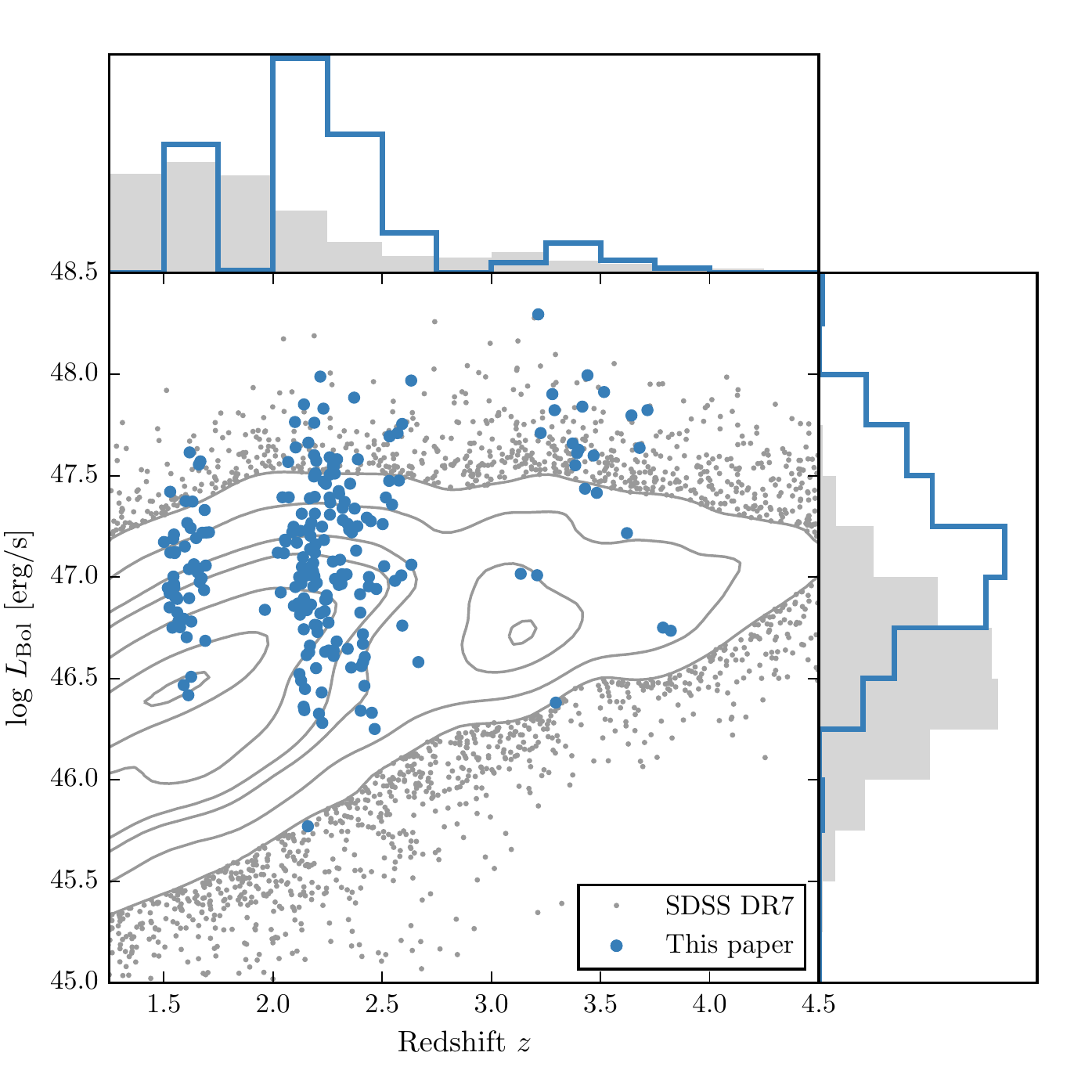} 
    \caption{The ranges in redshift and luminosity covered by our sample, relative to the redshift-luminosity distribution of the SDSS DR7 quasar catalogue. In regions of high point-density, contours show equally-spaced lines of constant probability density generated using a Gaussian kernel-density estimator. For the SDSS sample we use \citet{hewett10} redshifts and bolometric luminosities measured by \citet{shen11}. For the quasars in this paper the redshift is defined using the peak of the \hans/\hb emission and the luminosity is measured in the continuum at 1350\AA\, and converted to a bolometric quantity using the same conversion factor employed by \citet{shen11}.}     
    \label{fig:lzplane}
\end{figure}

In Fig.~\ref{fig:lzplane} we show the luminosities and redshifts of the quasar sample relative to the redshift-luminosity distribution for the Sloan Digital Sky Survey \citep[SDSS;][]{york00} Seventh Data Release \citep[DR7;][]{schneider10}.
Our sample spans a redshift range $1.5 < z < 4.0$ and a bolometric luminosity range $10^{45.5}-10^{48}$\,\ergs. 
Spectra were obtained within one or more of the $JHK$ pass-bands and the gaps in our sample coverage at $z\sim1.8$ and $z\sim3$ are due to the presence of atmospheric absorption. 
Obtaining near-infrared spectra of adequate resolution and signal-to-noise ratio (S/N) of even moderately bright quasars remains resource intensive. 
As a consequence, at fixed redshift, the luminosities of the quasars are brighter than the average luminosity of the SDSS sample, although the dynamic range in luminosity is a full 1.5 decades.

Below, we present the key elements of the observations of the six quasar sub-samples that make up the full 230-quasar catalogue.

\begin{table*}
  \centering
  \caption{The numbers of quasars with reliable \ha and \hb line measurements, the spectrographs and telescopes used to obtain the near-infrared spectra, and the instrumental configurations.}
  \label{tab:specnums}
  \begin{minipage}{16cm}
  \centering
    \begin{tabular}{cccccccc} 
    \hline
    Spectrograph & Telescope & Resolving power & Wavelength coverage & Slit width & Exposure times & \ha Sample & \hb Sample \\
    & & $\lambda/\Delta\lambda$ & $\mu$m & arcsec & hr & & \\  
    \hline
    FIRE       & MAGELLAN & 6000      & 0.80-2.50 & 0.6       & 0.5-1.0    & 18 & 19 \\
    GNIRS      & GEMINI-N & 5400      & 0.85-2.50 & 0.3-0.45 & 0.3-1.3  & 22 & 17 \\
    ISAAC      & VLT      & 5100      & 1.40-1.82 & 0.6       & 0.6-1.3  & 0  & 4 \\
    LIRIS      & WHT      & 945       & 1.39-2.42 & 1.0         & 0.2-0.8  & 15 & 0 \\
    NIRI       & GEMINI-N & 520-825   & 1.43-1.96 & 0.47-0.75 & 0.5-2.7 & 0  & 12 \\
    SINFONI    & VLT      & 2000-3000 & 1.10-2.45\footnote{$J$, $H$ or $K$ filters were employed to ensure coverage of the \hbns/\oiii spectral region.} &  & 0.1-0.7  & 2  & 25 \\
    SOFI       & NTT      & 1000-2000 & 1.53-2.52\footnote{Both the low resolution red grism and the medium resolution grism, with $K$ and $H$ filters, were employed.} & 0.6       & 0.5-1.8  & 47 & 23 \\
    TRIPLESPEC & ARC-3.5m & 2500-3500 & 0.95-2.46 & 1.1-1.5   & 1.0-1.5    & 33 & 20 \\
    TRIPLESPEC & P200     & 2500-2700 & 1.00-2.40   & 1.0         &          & 23 & 19 \\
    XSHOOTER   & VLT      & 4350-7450 & 0.30-2.50   & 0.5-1.6   &     0.2-0.8     & 4  & 7 \\
    \hline
    \multicolumn{6}{c}{Total} & 164 & 144 \\
    \hline
    \end{tabular}
    \end{minipage}
\end{table*}

\subsection{Near-infrared observations}

\subsubsection{Coatman et al. (2016) Quasars}
\label{ss:paperI}

\citet{coatman16} (hereafter Paper I) observed objects drawn from the SDSS DR7 quasar catalogue.
Quasars were selected to i) have redshifts $2.14 < z < 2.51$, ii) be radio-quiet, iii) show no evidence of broad absorption lines (BALs) affecting the \civ emission line, iv) be free from significant dust extinction and v) possess \civns-emission shapes spanning the full range in the population.
Near-infrared spectra, including the \ha line, were obtained with the Long-slit Intermediate Resolution Infrared Spectrograph \citep[LIRIS;][]{manchado98} mounted on the 4.2\,m William Herschel Telescope (WHT) at the Observatorio del Roque de los Muchachos (La Palma, Spain).
The spectra were reduced using standard \textsc{IRAF}\footnote{IRAF is distributed by the National Optical Astronomy Observatory, which is operated by the Association of Universities for Research in Astronomy (AURA) under a cooperative agreement with the National Science Foundation.} packages, as described in Paper I. 
We have selected 15 quasars with the highest S/N \ha spectra from the original sample of 19 (see Section~\ref{sec:flagged_spectra}) and the observational properties of these quasars are summarised in Table~\ref{tab:wht}. 

\subsubsection{Shen \& Liu (2012) and Shen (2016) Quasars}

\citet{shen16a} and \citet{shen12} obtained near-infrared spectroscopy for a sample of 74 luminous, $1.5 < z < 3.5$ quasars selected from the SDSS DR7 quasar catalogue. 
Targets had to possess good optical spectra covering the \civ line and have redshifts $z\sim$ 1.5, 2.1, and 3.3 to ensure that the \hbns-\oiii region was covered in one of the near-infrared $JHK$ bands.
Thirty-eight of the quasars were observed with TripleSpec \citep{wilson04} on the Astrophysics Research Consortium (ARC) 3.5\,m telescope, and 36 with the Folded-port InfraRed Echellette \citep[FIRE;][]{simcoe10} on the 6.5\,m Magellan-Baade telescope.
The reduction of the spectra is described in \citet{shen16a} and \citet{shen12}. 
The 57 quasars for which we were able to measure reliable emission line properties (Section~\ref{sec:flagged_spectra}) are summarised in Table~\ref{tab:shen}.

\subsubsection{Quasar Pairs}

Twenty per cent of our catalogue was observed as part of an ongoing effort to identify quasar pairs at very close projected separations \citep[Quasars Probing Quasars\footnote{www.ucolick.org/\textasciitilde xavier/QPQ/Quasars\_Probing\_Quasars} (QPQ);][]{hennawi06a,hennawi10}. 
The primary science driver of this work is to study the circum-galactic medium of the foreground quasars in absorption \citep{hennawi06b}.
Very accurate systemic redshift measurements are a requirement and a large amount of effort has gone into obtaining near-infrared spectra which cover low-ionisation broad lines or features from the quasar narrow line region \citep{prochaska09,lau15,hennawi15}. 
From the QPQ data set we identified 46 quasars with good-quality near-infrared spectra covering the \ha and/or \hb lines and SDSS and/or BOSS spectra covering the \civ line. 
Twenty-two quasars were observed with the Gemini Near-Infrared Spectrograph \citep[GNIRS;][]{elias06} on the 8.1 m Gemini North telescope, 4 using the Infrared Spectrometer And Array Camera \citep[ISAAC;][]{moorwood98b} on the European Southern Observatory (ESO) Very Large Telescope (VLT), 11 with the Near InfraRed Imager and Spectrometer \citep[NIRI;][]{hodapp03} also on Gemini North and 9 with XSHOOTER \citep{vernet11}, again, on the VLT. 
The broad wavelength coverage of XSHOOTER means that the spectra cover the region from \civ to \ha at the redshifts targeted. 
The XSHOOTER spectra have higher S/N and resolution than the SDSS/BOSS spectra in the rest-frame ultraviolet and therefore the XSHOOTER spectra are used by default to measure the \civ emission. 

The  XSHOOTER  spectra  were  reduced  with  a  custom  software  package  developed  by  George  Becker \citep[for details, see][]{lau15}. 
The remaining data was processed with algorithms in the LowRedux\footnote{www.ucolick.org/\textasciitilde xavier/LowRedux} package \citep[see][]{prochaska09}.

The 46 quasars for which we were able to measure reliable emission line properties are summarised in Table~\ref{tab:qpq}.

\subsubsection{VLT SINFONI Quasars}

We performed a search of the ESO archive for high-$z$ quasars observed with the SINFONI  integral  field  spectrograph \citep{eisenhauer03,bonnet04} at VLT/UT4.
We found 37 quasars with redshifts $1.5 < z < 3.7$ which have $H$ and/or $K$ SINFONI spectroscopy, covering the \hb and \ha lines respectively, where good optical spectroscopy covering \civ is also available. 
Thirty of the quasars are from a large programme led by L. Wisotzki (programme 083.B-0456(A)) to study the mass function and Eddington ratios of active BHs at redshifts $z\sim 2$ drawn from the Hamburg/ESO survey \citep{wisotzki00}.
The Hamburg/ESO optical spectra have a typical $\sim$400\kms\, spectral resolution and S/N $\gtrsim$ 10 per pixel. 
A further seven SINFONI spectra are from a programme led by  J. D. Kurk (programme 090.B-0674(B)) to obtain reliable BH mass estimates from \hans/\hb for a sample of radio-loud/radio-quiet SDSS quasars.

The SINFONI spectra were reduced using the package EASYSINF\footnote{www.mrao.cam.ac.uk/\textasciitilde rw480/easysinf}.  
The package, which is based on the ESO-SINFONI pipeline, is described in \citet{williams16}. 

The 25 quasars for which we were able to measure reliable emission line properties are summarised in Table~\ref{tab:sinf}. 

\subsubsection{ESO NTT SOFI Quasars}

Twelve per cent of the quasar catalogue derives from a large programme (programme 187.A-0645; PI: J. Hennawi) to combine near-infrared spectra from SOFI \citep{moorwood98a} on the 3.6\,m New Technology Telescope (NTT) with archival high-resolution optical spectra from the UV-Visual Echelle Spectrograph \citep[UVES;][]{dekker00} at VLT/UT2 and the High Resolution Echelle Spectrometer \citep[HIRES;][]{vogt94} at Keck to construct a legacy database of bright, high-redshift ($2 < z < 4$) quasars with both rest-frame optical spectra, covering the \hbns-\oiii complex, and high-resolution rest-frame ultraviolet spectra.
The main science goal is to obtain precise systemic redshifts which are crucial for the study of absorption line systems.  
The SOFI spectra were reduced using a custom data reduction pipeline using algorithms in the LowRedux package.

Eighteen quasars have been targeted as part of the SDSS/BOSS spectroscopic quasar surveys.
In addition, 13 reduced and fluxed UVES spectra have been made available to us by A. Dall'Aglio (a description of the reduction procedure is contained in \citet{dallaglio08}).
The sample is larger ($\sim$100 quasars) but reduced UVES spectra providing rest-frame ultra-violet coverage of \civ are not yet available for the remainder. 
The spectral resolution of the UVES observations is very high ($R$$\sim$40\,000) and the S/N of the spectra re-binned to a resolution of $\simeq$2000 is S/N$\simeq$300. 
The 28 quasars for which we were able to measure reliable emission line properties are summarised in Table~\ref{tab:sofijh}.

Over five nights from 2015 August 31 to September 4 we obtained near-infrared SOFI spectra for a further 26 quasars (programme 095.B-0644(A); PI: L. Coatman). 
These quasars were selected from the SDSS DR7 quasar catalogue using criteria very similar to those described in Paper I (see Section~\ref{ss:paperI}). 
In particular, we selected quasars with large \civ blueshifts to improve the statistics in this region of the \civ emission-line parameter space. 
The 27 quasars for which we were able to measure reliable emission line properties are summarised in Table~\ref{tab:sofilc}.

\subsubsection{P200 TripleSpec Quasars}

A further 36 quasars in our catalogue are bright SDSS quasars which were observed with the TRIPLESPEC spectrograph on the Palomar 200-inch Hale telescope (P200). 
The objects were observed with the same science goals as the SOFI NTT large programme. 
The spectra were reduced using a custom pipeline, again using algorithms in the LowRedux package. 
The 32 quasars for which we were able to measure reliable emission line properties are summarised in Table~\ref{tab:triple}.

\subsection{Optical observations}

In the previous sections, we described the infrared spectra of the 230 quasars making up our full spectroscopic catalogue. 
We will now describe the companion optical spectra, which provide coverage of the \civ emission. 

Optical SDSS DR7 spectra are employed for 70 quasars in the full catalogue.  
The SDSS DR7 spectra are moderate resolution ($R$$\simeq$2000) and S/N (S/N$\simeq$20) and cover the observed-frame wavelength interval $\sim3800-9180$\,\AA.
Many of the quasars in the SDSS DR7 catalogue have been re-observed as part of the Sloan Digital Sky Survey-III: Baryon Oscillation Spectroscopic Survey \citep[SDSS-III/BOSS;][]{dawson13}. 
As the BOSS-spectra typically have higher S/N than the SDSS DR7 spectra, we have used the BOSS spectra when available (126 quasars).  
We also use high-resolution optical spectra taken with VLT/UVES (11 quasars) and VLT/XSHOOTER (8 quasars), and Hamburg/ESO spectra for a further 15 quasars.  

In summary, we have assembled a sample of 230 luminous, high-$z$ quasars with optical and near-infrared spectra.
This will allow us to directly compare virial BH mass estimates based on the \civ line-width with estimates based on the line-widths of the low-ionisation Balmer lines \ha and \hbns.  

\section{Spectral Measurements}

Conventionally, single-epoch virial estimates of the BH mass are a function of the line-of-sight velocity width of a broad emission line and the quasar luminosity. 
The velocity width is a proxy for the virial velocity in the broad line region (BLR) and, as revealed in reverberation-mapping studies, the luminosity is a proxy for the typical size of the BLR \citep[the $R-L$ relation; e.g.][]{kaspi00,kaspi07}. 
Most reverberation mapping campaigns have employed \hb time-lags and velocity widths, but the line-widths of \ha and \mgiins$\lambda$2800 have been shown to yield consistent BH masses \citep[e.g.][]{mclure02,greene05,onken08,shen08,wang09,rafiee11,mejia-restrepo16}. 
In Section~\ref{sec:hahbcomparison} we verify that this is the case for the 99 quasars in our sample with measurements of both \ha and \hb lines.     

At redshifts $z> 2.2$, where the hydrogen Balmer lines and \mgii are no longer accessible in many optical spectra, the \civns$\lambda$1550 emission doublet has routinely been used to provide estimates of the virial velocity \citep[e.g.][]{shen11}. 
As has long been recognised \citep{gaskell82, tytler92} the \civ emission line in many quasars includes contributions from gas that does not straightforwardly relate to virial motions within a stable BLR.
A number of studies \citep[e.g.][]{shen08, richards11} have shown that the amplitude of the systematic shift of the \civ emission to shorter wavelengths (relative to the systemic velocity) is strongly correlated with the properties of the emission-lines and the overall spectral energy distributions (SEDs). 

In our work, a robust measure of the \civ emission-line `blueshift' provides the basis for the corrected \civ velocity-width measurements, and hence BH masses.
The effectiveness of the scheme is validated via a direct comparison of the \civ velocity-widths to the Balmer emission velocity-widths in the same quasars. 
Our process is as follows. 
First, an accurate measure of the quasar's systemic redshift is required, for which we adopt the centre of the Balmer emission, where the centre, $\lambda_{\rm half}$, is the wavelength that bisects the cumulative total flux. 
Balmer emission centroids are available for all quasars in the catalogue but we verify that the measure is relatively unbiased through a comparison of the centroids to the wavelengths of the peak of the narrow \oiiins\ll4960,5008 doublet for the subset of spectra where both are available. 
Second, the blueshift of the \civ emission line is determined. 
Again, we adopt the line centroid to provide a robust measure of the \civ emission blueshift.
The blueshift (in \kms) is defined as $c\times$(1549.48-$\lambda_{half}$)/1549.48 where $c$ is the velocity of light and 1549.48\,\AA \ is the rest-frame wavelength for the \civ doublet, assuming equal contribution from both components.
Positive blueshift values indicate an excess of emitting material moving towards the observer and hence out-flowing from the quasar. 

Emission-line velocity widths are derived from the full-width-at-half-maximum (FWHM) of the lines but we also compute the line dispersion (calculated from the flux-weighted second moment of the velocity distribution) as some authors have claimed this provides a better estimate of the virial velocity \citep{denney13}. 

To minimise the impact of the finite S/N of the quasar spectra and the presence of absorption features superposed on the broad emission lines we first fit a parametric model to the continuum and the emission lines. 
The purpose of the parametric fits is, however, simply to provide higher S/N representations of the emission features. 
The particular form of the model parametrizations is not important and the fits are used only to provide robust line parameters, such as the centroid $\lambda_{\rm half}$, and FWHM, which are measured non-parameterically from the best-fitting model. 
The models used and the fitting procedure are described below. 
The issues involved in deriving parameters for broad emission lines from spectra of modest S/N -- for example, subtraction of narrow line emission, subtraction of \feii emission -- have been covered comprehensively by other authors \citep[e.g.][]{shen11,shen12,denney13,shen16a} and, as far as possible, we follow standard procedures described in the literature. 

\subsection{\civns}
\label{sec:civ}

The parametrization of the \civ emission line is identical to the one described in Paper I. 
We first define a power-law continuum, $f(\lambda) \propto \lambda^{-\alpha}$, with the slope, $\alpha$, determined using the median values of the flux in two continuum windows at 1445-1465 and 1700-1705\AA. 
The continuum emission is subtracted from the spectra, which is then transformed from wavelength units into units of velocity relative to the rest-frame line-transition wavelength for the \civ doublet (1549.98\AA, assuming equal contributions from both components). 
The parametric model is ordinarily fit within the wavelength interval 1500-1600\AA\, (corresponding to approximately $\pm 10\,000$ \kms\, from the rest-frame transition wavelength), a recipe that is commonly adopted \citep[e.g.][]{denney13}. 
The line-window was extended if more than 5\,per cent of the total flux in the profile was present blueward of the short wavelength limit. 
Narrow absorption features, which are frequently found superimposed on \civ emission (see, for example, the \civ profile of J0942+3523 in Fig.~\ref{fig:examplegrid}), were masked out during the fit.

The \civ emission was fit with sixth-order Gauss-Hermite (GH) polynomials, using the normalisation of \citet{marel93} and the functional forms of \citet{cappellari02}. 
We allowed up to six components, but in many cases a lower order was sufficient (40 and 45 per cent were fit with second- and fourth-order GH polynomials respectively).
GH polynomials were chosen because they are flexible enough to model the often very asymmetric \civ line profile. 
The flip-side of this flexibility, however, is that the model has a tendency to over-fit when spectra possess low S/N. 
The fits were therefore carefully checked visually and the number of components reduced if over-fitting was evident.

In Paper I we found that using the commonly employed three-Gaussian component model, rather than the GH polynomials, resulted in only marginal differences in the line parameters. 
Our best-fit parameters are also in good agreement with \citet{shen11}, who employ a multi-Gaussian parametrization. 
In the Appendix we demonstrate that the mean difference between our FWHM measurements and the measurements of \citet{shen11} is just 200\kms for the quasars common to both samples, which is much too small to have any significant effect on our results.

\subsection{\ha}

A power-law continuum is fit using two continuum windows at 6000-6250 and 6800-7000\,\AA. 
The continuum-subtracted flux is then fit in the wavelength interval 6400-6800\,\AA. 
We adopt a rest-frame transition wavelength of 6564.89\,\AA\, to transform wavelengths into equivalent Doppler velocities. 
The broad component of \ha is fit using one or two Gaussians, constrained to have a minimum FWHM of 1200\kms. When two Gaussians are used, the velocity centroids are constrained to be the same.

The emission-line profiles of both \hb and \ha frequently include a significant narrow component from the physically more extended narrow line region (NLR). 
Additional Gaussian components were included in our parametric model to fit the narrow component of \ha as well as [\niins]\ll6548,6584 and [\siins]\ll6717,6731.
This resulted in a better fit to the observed flux in 50\, per cent of cases. 
We impose a 1200\kms upper limit on the FWHM of all narrow lines and the amplitudes of all components must be non-negative.
The relative flux ratio of the two [\niins] components is also fixed at the expected value of 2.96.
In 70\, per cent of the spectra the \oiiins\ll4960,5008 doublet is detected at moderate S/N in the \hb region. 
In these cases the peak of the \oiii is used to fix the velocity offsets and the FWHMs of the narrow line components in the \ha region.  
For spectra where the \oiii doublet does not constrain the velocity and FWHM accurately, the narrow emission in the \ha and \hb regions are fitted independently but, for each region, the individual narrow-line velocity offsets and the FWHMs are constrained to be identical. 
In these objects the narrow line contribution is generally weak, and so does not have a large effect on the line parameters we measure for the broad component.   

The model described above is very similar to the one described in \citet{shen16a}, \citet{shen12} and \citet{shen11}, the only major difference being that we do not fit the \ha and \hb emission regions simultaneously. 
In Appendix~\ref{appendix:shen}, we compare our \ha line measurements for the subset of our sample taken from \citet{shen16a} and \citet{shen12}, and find a scatter of just 0.07 dex. 

\subsection{\hb and \oiiins}

Emission from optical \feii is generally strong in the vicinity of \hbns.
We therefore fit a combination of a power-law continuum and an optical \feii template -- taken from \citet{boroson92} -- to two windows at 4435-4700 and 5100-5535\,\AA. 
The \feii template is convolved with a Gaussian, and the width of this Gaussian, along with the normalisation and velocity offset of the \feii template, are free variables in the pseudo-continuum fit.
We use the same model to fit the broad and narrow components of \hb as was used with \hans. 
Each line in the \oiii doublet is fit with two Gaussians, to model both the systemic and any outflow contributions. 
The peak flux ratio of the \oiii 4960\,\AA\, and 5008\AA\, lines is fixed at 1:3. 
As for the fit to the narrow lines in the spectral region around \hans, the width and velocity offsets of all the narrow components are set to be equal, and an upper limit of 1200\kms is placed on the FWHM. 

\subsection{Fitting procedure}

Model parameters were derived using a standard variance-weighted least-squares minimisation procedure employing the Levenberg-Marquardt algorithm. 
Prior to the fit, the spectra were inspected visually and regions significantly affected by absorption or of low S/N were masked out.

\begin{figure*}
    \includegraphics[width=2\columnwidth]{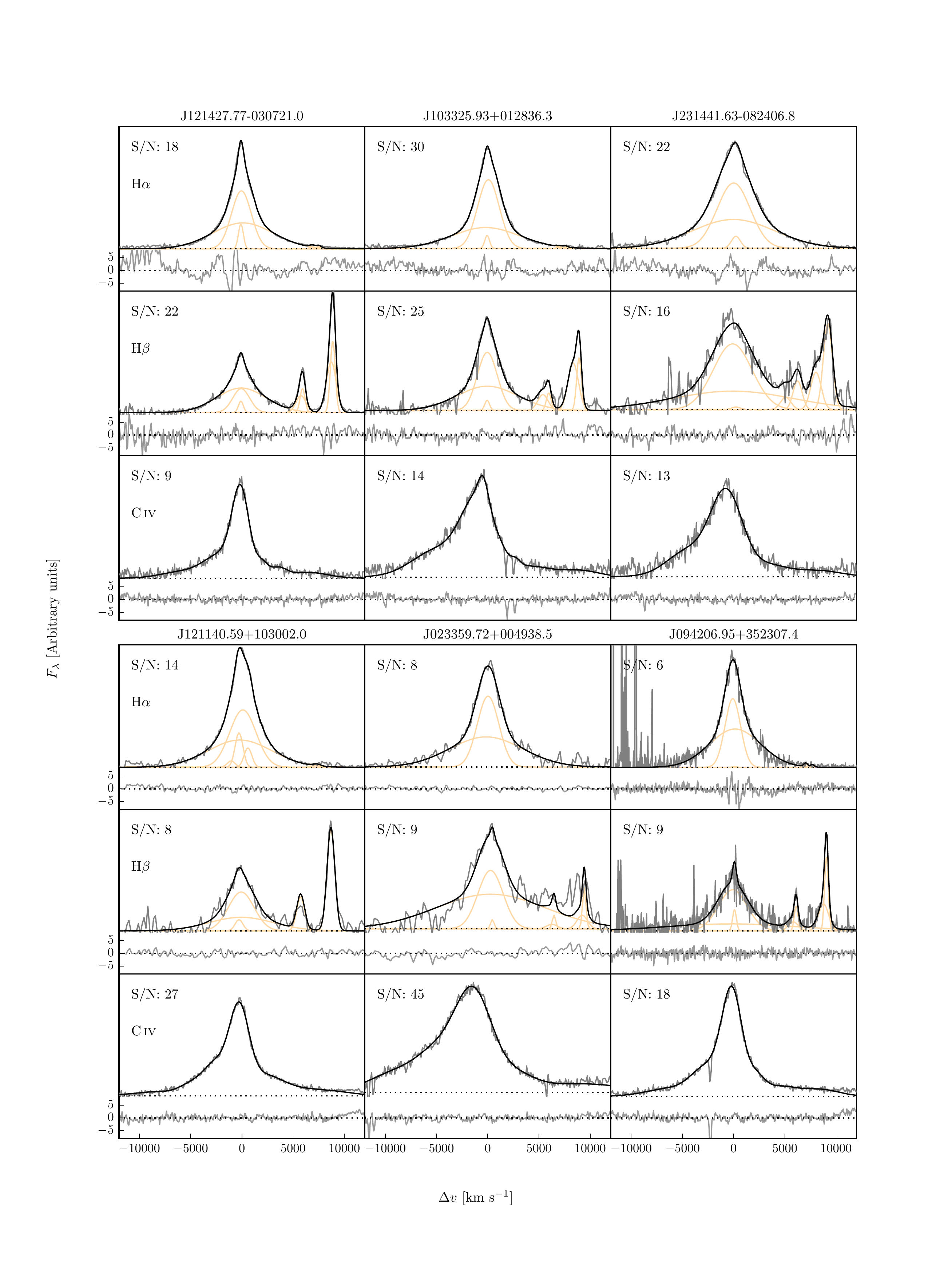} 
    \caption{Model fits to continuum-subtracted \hans, \hbns, and \civ emission in six quasars, chosen to represent the range of S/N (indicated in the figure and given per 150\kms\, pixel in the continuum) and line shapes present in the catalogue. The data is shown in grey, the best-fitting parametric model in black, and the individual model components in orange. The centroid of the broad \ha emission is used to set the redshift, and $\Delta{v}$ is the velocity shift from the line rest-frame transition wavelength. Below each fit we plot the data minus model residuals, scaled by the errors on the fluxes.} 
    \label{fig:examplegrid}
\end{figure*}

In Fig.~\ref{fig:examplegrid} we present our parametric fits to the \civ, \ha and \hb emission lines in a handful of quasars, which have been chosen to illustrate the range of spectrum S/N and line shapes in the sample.  
The mean reduced chi-squared values in our \hans, \hb and \civ fits are 1.69, 1.62, and 1.77 respectively and, in general, there are no strong features observable in the spectrum minus model residuals. 
Table~\ref{tab:specmeasure} includes the line parameters of our best-fitting model for each line.
The reported line-width measures are corrected for instrumental broadening by subtracting the resolution of the spectrograph in quadrature. 
The spectrograph resolutions, which we estimate from the line widths in the observed sky spectra, range from 25\kms\, for XSHOOTER to 477\kms\, for the low-resolution LIRIS grism and are therefore small relative to the quasar broad line widths.

\begin{table*}
  \centering
  \vspace*{-0.4cm}
  \caption{The format of the table containing the emission line properties from our parametric model fits. The table is available in machine-readable form in the online journal.}
  \label{tab:specmeasure}
  \vspace*{-0.1cm}
  \begin{minipage}{16cm}
  \centering
    \begin{tabular}{ccc} 
    \hline
    & Units & Description \\ 
    \hline
    NAME & & Catalogue name \\
    FWHM\_BROAD\_HA & \kms & FWHM of broad \ha line \\ 
    FWHM\_BROAD\_HA\_ERR & \kms & \\
    SIGMA\_BROAD\_HA & \kms & Dispersion of broad \ha line\\
    SIGMA\_BROAD\_HA\_ERR & \kms & \\
    Z\_BROAD\_HA & & Redshift from broad \ha line\\
    FWHM\_BROAD\_HB & \kms & FWHM of broad \hb line \\
    FWHM\_BROAD\_HB\_ERR & \kms & \\
    SIGMA\_BROAD\_HB & \kms & Dispersion of broad \hb line \\
    SIGMA\_BROAD\_HB\_ERR & \kms & \\
    Z\_BROAD\_HB & & Redshift from broad \hb line\\
    FWHM\_CIV & \kms & FWHM of \civ doublet \\
    FWHM\_CIV\_ERR & \kms & \\
    SIGMA\_CIV & \kms & Dispersion of \civ doublet \\
    SIGMA\_CIV\_ERR & \kms & \\
    BLUESHIFT\_CIV\_HA & \kms & Blueshift of \civ relative to centroid of broad \hans \\
    BLUESHIFT\_CIV\_HA\_ERR & \kms & \\
    BLUESHIFT\_CIV\_HB & \kms & Blueshift of \civ relative to centroid of broad \hbns \\
    BLUESHIFT\_CIV\_HB\_ERR & \kms & \\
    LOGL5100 & \ergs & Monochromatic continuum luminosity at 5100\AA \\
    LOGL1350 & \ergs & Monochromatic continuum luminosity at 1350\AA\\
    \hline
    \end{tabular}
  \vspace*{-0.4cm}
  \end{minipage}
\end{table*}

\subsection{Spectra removed from sample}
\label{sec:flagged_spectra}

\begin{table}
  \centering
  \caption{The number of spectra removed from our sample by the cuts described in Section~\ref{sec:flagged_spectra}.}
  \label{tab:flagged_spectra}
    \begin{tabular}{cccc}
    \hline
    & & \ha sample & \hb sample \\ 
    \hline
    \multicolumn{2}{c}{Total} & 194 & 279 \\
    \hline
    \hans/\hbns & Wavelength & 6 & 27 \\
    & S/N & 8 & 83 \\
    \hline
    \civ & Wavelength & 6 & 5 \\
    & S/N & 4 & 12 \\
    & Absorption & 6 & 8 \\
    \hline
    \multicolumn{2}{c}{Total remaining} & 164 & 144 \\
    \hline
    \end{tabular}
\end{table}

Through visual inspection we flagged and discarded the spectra of quasars for which reliable emission line parameters could not be obtained.

First, we flagged emission lines in spectra that possessed insufficient S/N. 
A single minimum S/N threshold was not entirely effective and, instead, spectra were flagged when it was judged conservatively that no meaningful constraints could be placed on the velocity centroid and/or width of the emission-line. 

Second, we flagged emission lines where significant regions of the continuum and/or emission line fell outside of the wavelength coverage of the spectra. 
Reliable continuum definition and subtraction is not straightforward for emission lines so affected. 

Third, we flagged \civ emission lines because of strong, narrow absorption close to the peak of the line where reliable interpolation across the absorption, using our parametric model, was not possible. 

\begin{figure}
    \includegraphics[width=\columnwidth]{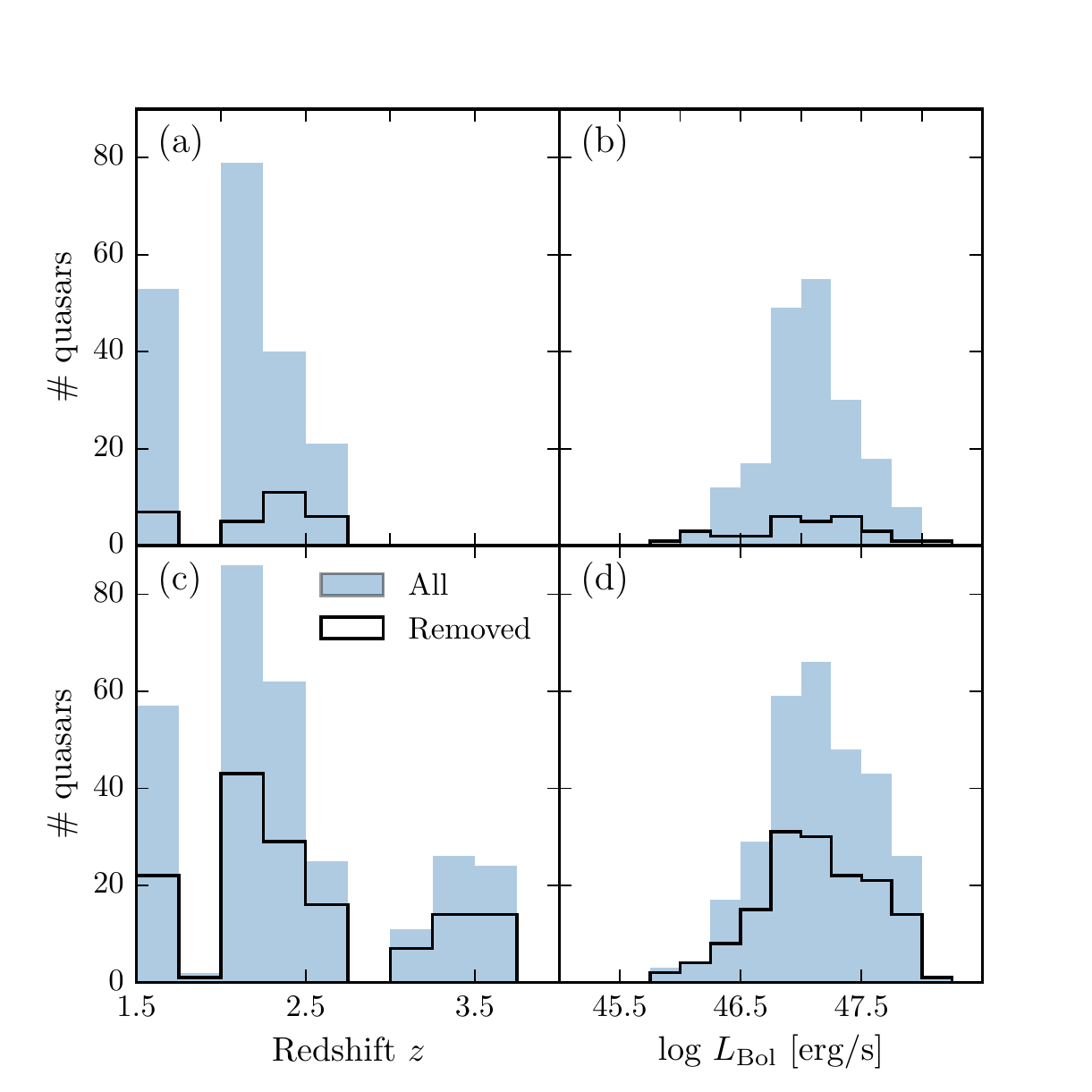} 
    \caption{The redshift and luminosity distributions of the spectra removed from our \hans/\civ (a, b) and \hbns/\civ (c, d) samples.} 
    \label{fig:flagged_spectra}
\end{figure}

The number of spectra that are removed by each cut is given in Table~\ref{tab:flagged_spectra} and the distribution in redshift and luminosity is shown in Fig.~\ref{fig:flagged_spectra}. 
Unsurprisingly, there is a preferential removal of intrinsically faint quasars, whose spectra can be of poorer S/N, and a loss of quasars at redshifts $z\sim2.6$ where the \ha emission falls at the edge of the $K$-band.
\hb is much weaker than \hans, and the \hb spectra are generally of lower S/N. 
As a result, the fraction of \hb spectra that are flagged -- 39 per cent -- is particularly high.   

\subsection{Emission-line parameter uncertainties}

The 1$\sigma$ error bars calculated from the covariance matrix in least-squares minimisation will underestimate the true uncertainties on the line parameters, since they do not account for systematic errors such as the significant uncertainty introduced in the continuum subtraction procedure.  
To calculate more realistic uncertainties on our fitted variables we employed a Monte Carlo approach. 
One thousand artificial spectra were synthesised, with the flux at each wavelength drawn from a Normal distribution (mean equal to the measured flux and standard deviation equal to the known error).
Our emission-line fitting recipe was then implemented on each of these mock spectra. 
The uncertainty in each parameter is given by the spread in the best-fitting values from the one thousand realisations of the fitting routine. 
In some cases the standard deviation of the parameter distribution was biased by extreme values caused by bad fits\footnote{In the analysis of the real spectra such fits are identified via visual inspection.}. 
We therefore chose to measure the spread in the parameter distribution by fitting a composite model with two Gaussian components -- one to model uncertainty in the parameter and the other any possible outlier component. 
The uncertainty in each line parameter was then taken to be the width of the narrower Gaussian. 

\subsection{Contemporaneity of spectra}

The epochs of the near-infrared and optical spectra can differ by many years.
For example, the NTT SOFI spectra were taken $\sim$14 years after the SDSS spectra, and the VLT SINFONI spectra 20 years or more after the Hamburg/ESO observations\footnote{Time differences in the quasar rest-frame are reduced by a factor of ($1 + z$).}.
If the broad emission line profiles varied significantly on these time-scales the relation between the \civ and Balmer line-width measurements could be blurred. 

Cases do exist of dramatic changes in quasar spectra over short time-scales, but this phenomenon is rare \citep{macleod16}. 
In our spectroscopic catalogue there are 112 SDSS DR7 quasars which are re-observed in BOSS and included in the DR12 quasar catalogue. 
The mean time elapsed between the two sets of observations is $\sim$8 years. 
The root-mean-square difference in the \civ FWHM measured from the BOSS and SDSS spectra is a modest $\simeq$500\kms. 
Differences in the S/N of the spectra will make a substantial contribution and the scatter due to true variations in the \civ velocity-width will be significantly smaller than 500\kms. 
We conclude therefore that any intrinsic changes with time do not materially affect the emission line measurements.

\subsection{Quasar monochromatic luminosity}

Computing virial BH masses also requires the quasar luminosity in an emission-line free region of the continuum adjacent to the broad line being used. 
The luminosity is used as a proxy for the size of the BLR. 
The monochromatic continuum flux is generally measured at 1350\,\AA\ for \civ and 5100\,\AA\, for \ha and \hbns. 

Relative flux-calibration of the infrared spectra as a function of wavelength has been achieved, to $\simeq$10 per cent, through observations of appropriate flux standards. 
The absolute flux levels, however, can be in error by large factors due to variable atmospheric conditions combined with the narrow slit widths. 
For the majority of the quasars we have, therefore, established the absolute flux scale for each near-infrared spectrum using the same quasar SED-model fitting scheme employed in Paper I.
The SED model, described in \citet{maddox12}, gives a very good fit to the SDSS and UKIDSS magnitudes of SDSS DR7 quasars, reproducing the individual magnitudes with a $\sigma <$0.1\,mag. 
For 207 quasars, ($Y$)$JHK$ passband magnitudes from the UKIRT Infrared Deep Sky Survey \citep[UKIDSS;][]{lawrence07} Large Area Survey, the Two Micron All Sky Survey \citep[2MASS;][]{skrutskie06} and the Visible and Infrared Survey Telescope for Astronomy (VISTA) Hemisphere Survey \citep[VHS;][]{mcmahon13} and Kilo-Degree Infrared Galaxy \citep[VIKING;][]{edge13} survey are available. 
The SED model was fit to the infrared magnitudes; integrating the SED model through the pass-band transmission functions, to give model magnitudes, and performing a variance weighted least-squares fit to the observed magnitudes. 
The flux at 5100\,\AA\, was then taken from the normalised model.

For 19 of the remaining 23 quasars, where near-infrared photometry was not available, the quasar SED model was fit to the SDSS spectra, the flux calibration of which are known to be excellent.  
The fit was done using a simple variance-weighted chi-squared minimisation procedure in emission line-free intervals of the optical spectra.   
The model includes a reddening, $E(B-V)$, based on a Small Magellanic Cloud-like extinction curve, and an overall normalisation of the model as free parameters.
In practice, the quasars possess only very modest reddenings, with $E(B-V)\simeq$0.0-0.1.
The flux at 5100\,\AA\, was then, again, taken from the normalised SED model.
For the four remaining quasars, which possess neither near-infrared photometry nor SDSS DR7 spectra, we fit the SED model to the BOSS DR12 spectra. 
To avoid the known issues in the flux calibration of the BOSS DR12 quasar spectra at observed-frame blue wavelengths \citep{lee13}, our fitting was confined to rest-frame wavelengths long-ward of 1275\AA. 

Comparison of the 5100\,\AA\, luminosity, computed using the photometry- and spectrum-based methods for 177 quasars, showed a scatter of just $\sim$0.1\,dex.
We therefore assume 0.1\,dex to be the measurement uncertainty on the 5100\,\AA\, luminosities. 

For 34 quasars in the catalogue the optical spectra come from surveys other than SDSS/BOSS and optical magnitudes from recent epochs are not available. 
In order to obtain an estimate of the luminosity at 1350\,\AA\, for the 30 quasars, we take the standard \citet{maddox12} quasar SED model, normalised to the near-infrared magnitudes, and read off the flux at 1350\,\AA. 

For all the catalogue quasars, the optical and near-infrared spectra as well as the near-infrared photometry were obtained at different epochs, with rest-frame time differences of up to $\sim$5\,years. 
Intrinsic quasar photometric variability in the rest-frame ultraviolet and optical will therefore add additional scatter of $\sim$0.2\,mag \citep[e.g.][]{macleod10} to the derived 1350- and 5100\,\AA-luminosities. 

Given that the luminosity enters into the calculation of BH-mass only as the square-root, the uncertainty on the luminosities does not make a large contribution to the uncertainties in the BH mass estimates.  

\section{An empirical correction to \civ-based virial BH-mass estimates}

\subsection{\hans/\hb FWHM comparison}
\label{sec:hahbcomparison}

BH-mass calibrations which use the width of the broad \hb emission line as a proxy for the virial velocity are widely regarded as the most reliable, since most reverberation mapping employs the \hb line and the $R-L$ relation has been established using \hbns.
When \hb is not available, \ha has been shown to be a reliable substitute \citep[e.g.][]{greene05,shen11,shen12}. 

\begin{figure}
    \includegraphics[width=\columnwidth]{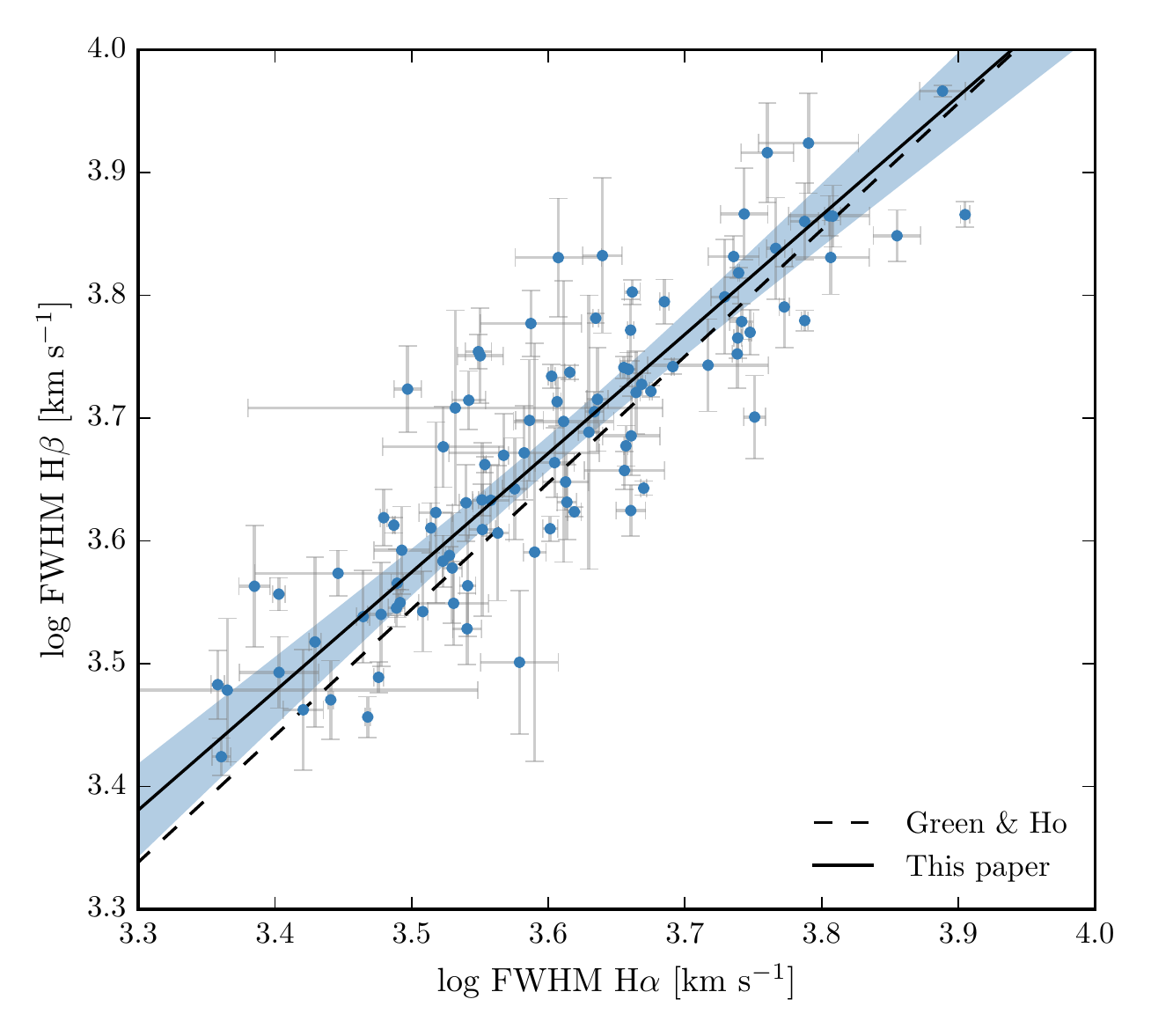} 
    \caption{Comparison of \ha and \hb FWHM measurements for 99 quasars. The solid line is our best-fitting power-law model, and the blue-shaded region shows the 2-$\sigma$ uncertainties on the model parameters. The dashed line is the relation found by \citet{greene05} using a sample of $z<0.35$ SDSS AGN.} 
    \label{fig:hahbcomp}
\end{figure}

In our sample, we have 99 quasars with reliable measurements of both \ha and \hb lines. 
The 99 objects include 21 quasars which were excluded from the main 308-object catalogue because adequate measurements of the \civ FWHM and blueshift could not be acquired. 
The line widths are compared in Fig.~\ref{fig:hahbcomp} and, as expected, a tight correlation is observed.  
\citet{greene05}, using a sample of 162 quasars with high S/N SDSS spectra at $z < 0.35$, established the following relation between the \ha and \hb FWHMs

\begin{equation}
  \rm{FWHM}(\rm{H}\beta) = \left( 1.07 \pm 0.07 \right) \times 10^3 \left( \frac{ \rm{FWHM}(\rm{H}\alpha) }{10^3 ~\rm{km}~\rm{s}^{-1} } \right)^{(1.03 \pm 0.03)}
\end{equation}

The relation is shown as the dashed line in Fig.~\ref{fig:hahbcomp}.
The root-mean-square scatter about this relation is 0.07\,dex, compared to the $\sim$0.1\,dex found by \citet{greene05}. 
However, we find a systematic offset, in the sense that the \hb line-widths we measure are on average larger by 270\kms\, than predicted by the \citet{greene05} relation. 
As our sample covers higher redshifts and luminosities than the sample in \citet{greene05}, we derive a new relation between the \ha and \hb FWHMs.       

We assume a relation of the same form used by \citet{greene05}, i.e. a simple power-law, and infer the model parameters by fitting a linear model (with slope $\alpha$ and intercept $\beta$) in log-log space.
The fit is performed within a Bayesian framework described by \citet{hogg10}. 
Each data point is treated as being drawn from a distribution function that is a convolution of the projection of the point's covariance tensor, of variance $\Sigma_i^2$, with a Gaussian of variance $V$ representing the intrinsic variance in the data.
The log-likelihood is then given by 

\begin{equation}
  {\rm ln} {\cal L} = - \sum_{i=1}^N \frac{1}{2} {\rm ln}\left[2\pi\left(\Sigma_i^2 + V\right)\right] - \sum_{i=1}^N \frac{\Delta_i^2}{2[\Sigma_i^2 + V]} 
\end{equation}

\noindent where $\Delta_i$ is the orthogonal displacement of each data point from the linear relationship. 
An advantage of this approach is that it allows a proper treatment of the measurement errors on both variables, which in this case are comparably large.
The model also makes the reasonable assumption that there is an intrinsic scatter in the relationship between the variables that is independent of the measurement errors.  
Following the suggestion by \citet{hogg10}, the linear model was parametrized in terms of ($\theta$,~$b_\bot$), where $\theta$ is the angle the line makes with the horizontal axis and $b_\bot$ is the perpendicular distance from the line to the origin.
Uniform priors were placed on these parameters, and the Jeffreys prior (the inverse variance) was placed on the intrinsic variance. 
The posterior distribution was sampled using a Markov Chain Monte Carlo (MCMC) method using the Python package {\tt emcee} \citep{foreman13}. 
 
\begin{figure}
    \includegraphics[width=\columnwidth]{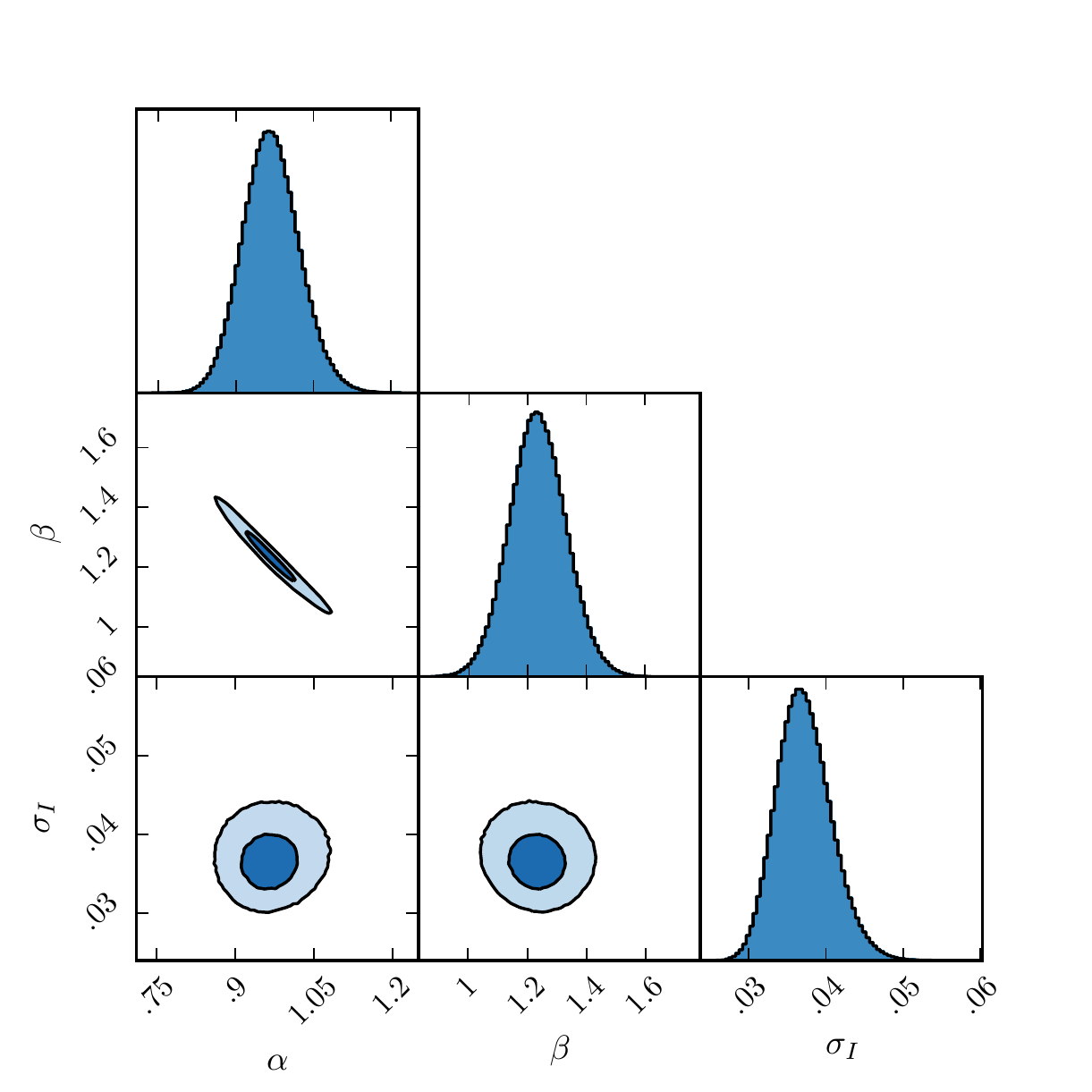} 
    \caption{One- and two-dimensional projections of the MCMC sampling of the posterior distribution from the fit in Fig.~\ref{fig:hahbcomp}. $\alpha$ is the power-law index, $10^\beta$ is the normalisation, and $\sigma_I$ is the intrinsic scatter. In the two-dimensional projections, 1- and 2-$\sigma$ contours are shown.} 
    \label{fig:ha_hb_mcmc_samples}
\end{figure}

The one- and two-dimensional posterior distributions are shown in Fig.~\ref{fig:ha_hb_mcmc_samples}. 
The solid line in Fig.~\ref{fig:hahbcomp} is the maximum likelihood solution

\begin{equation}
  \label{eq:ha2hb}
  \rm{FWHM}(\rm{H}\beta) = \left( 1.23 \pm 0.10 \right) \times 10^3 \left( \frac{\rm{FWHM}(\rm{H}\alpha)}{10^3 {\rm km\,s^{-1}}} \right)^{0.97 \pm 0.05}
\end{equation}

\noindent and the shaded region shows the 2$\sigma$ uncertainties on the model parameters.

As discussed above, our relation is displaced to slightly higher \hb FWHM than the \citet{greene05} relation -- the offset is 210\kms\, for a quasar with \ha FWHM 4500\kms.  
We infer a power-law index that, although slightly shallower, is consistent with the \citet{greene05} index within the quoted uncertainties. 
The intrinsic scatter in the data, $\sigma_I$, we infer from  the fit is 0.04 dex. 
This is smaller than the total scatter seen in Fig.~\ref{fig:hahbcomp} (0.06 dex), which suggests that measurement errors make a significant contribution to the total scatter in the relation. 

For 19 of the 99 quasars with \hb and \ha emission profiles, one of the two Gaussians used to reproduce the \hb profiles has a FWHM greater than 20\,000\kms and a fractional contribution to the total \hb broad line flux of $>$0.3 \citep{marziani09,marziani13}.  
Such a broad component is not seen in the \ha profiles and the very broad \hbns-component may be an artifact of the fitting scheme.
A particular issue for \hb is the presence of \feii emission, often at a significant level.
Furthermore, additional lines could be contributing to the underlying continuum \citep[e.g. the \heins\ll4922,5017 doublet;][]{veron02,zamfir10}. 

In Sec.~\ref{sec:correction} we use the whole of the \hb profile to derive an un-biased BH mass.  
If, instead, the FWHM is calculated from the narrower of the two Gaussian components rather than the composite profile, then the \hb FWHM decreases by 630\kms on average.
The \ha FWHM, which are calibrated against the \hb FWHM, will also decrease by the same amount on average. 
This will enhance the \civ FWHM relative to the \hans/\hb FHWM by $\sim$15\,per cent and increase the size of the correction which must be applied to the \civns-based BH masses by $\sim$30\,per cent.  

\subsection{Measuring the quasar systemic redshift}

An accurate measure of the quasar's systemic redshift is required in order for the blueshift of the \civ emission line to be determined.
Balmer emission centroids, where the centroid, $\lambda_{\rm half}$, is the wavelength that bisects the cumulative total flux, are available for all quasars in the catalogue and so we use this to define the systemic redshift.

For 83 and 120 quasars in the \ha and \hb samples respectively narrow \oiii emission is also detected. 
In the model fit to the \hb region the velocity centroids of the broad \hbns-line and the core component of the [\oiiins] emission were deliberately determined separately.
We find the intrinsic difference in the velocity centroids of the Balmer broad emission and the narrow \oiii emission to have a dispersion of 360\kms, which is very similar to the value found by \citet{shen16b}. 
However, the median velocity centroid of the narrow component of the \oiii emission is blueshifted by 270\kms\, relative to the centroid of the broad Balmer line. 
Applying our parametric model fitting routine to the composite spectrum from \citet{hewett10}, which is constructed using relatively low redshift SDSS quasars with $L_{\rm Bol}\sim10^{44}$\,erg\,s$^{-1}$, the centroids of the broad component of \hb and the narrow component of \oiii are found to be at essentially identical velocities, suggesting that the blueshifting of narrow \oiii could be luminosity dependent.
Regardless, since both the systematic offset and the scatter are small in comparison to the dynamic range in \civ blueshifts, the blueshift-based empirical correction we will derive does not depend on whether the broad Balmer emission or the \oiii centroid is used to define the systemic redshift. 

\subsection{Balmer/\civ line widths as a function of \civns-blueshift}
\label{sec:correction}

\begin{figure*}
    \includegraphics[width=2\columnwidth]{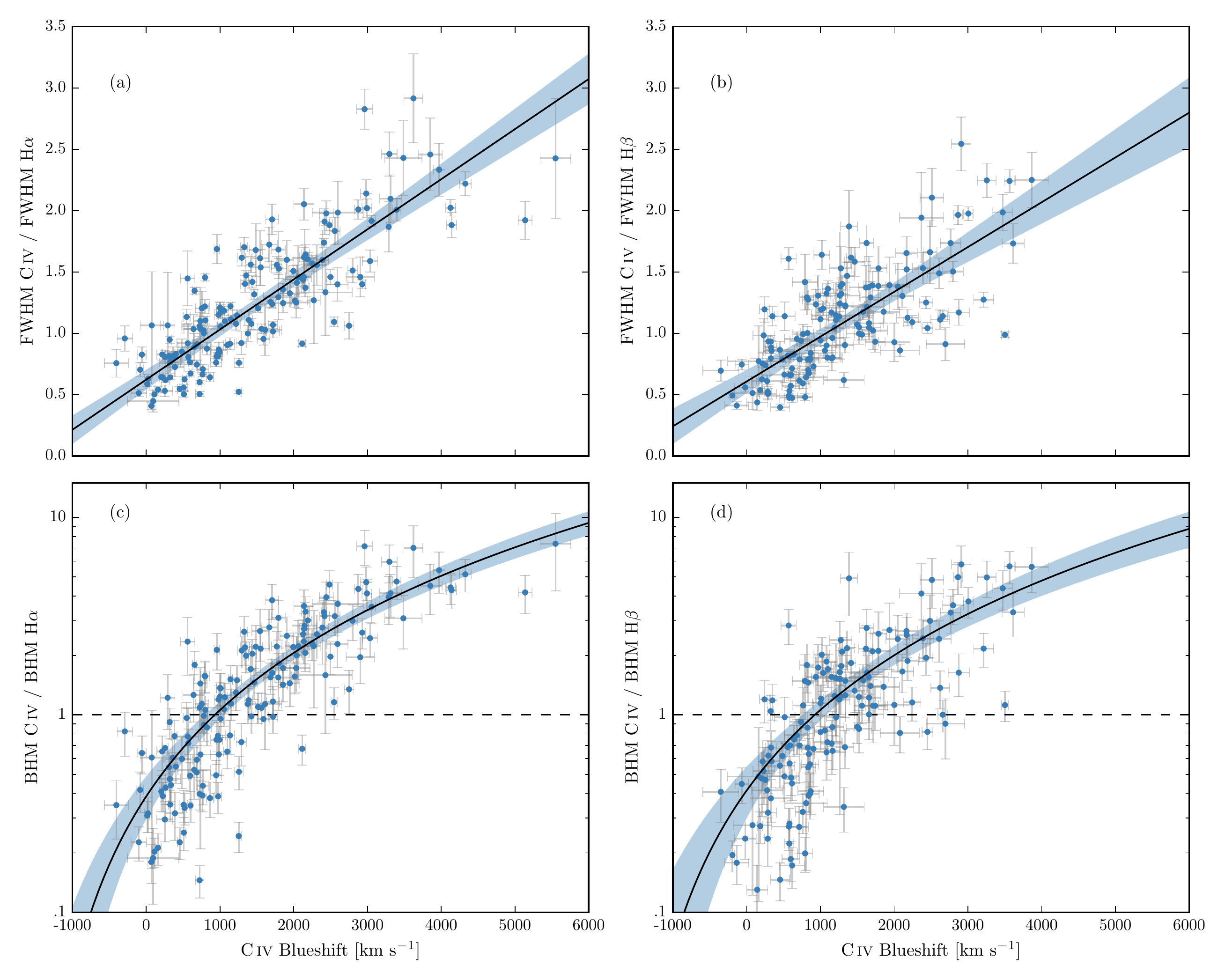} 
    \caption{\civ FWHM relative to \ha FWHM (a) and \hb FWHM (b), and \civ based BH mass (BHM) compared to \ha based mass (c) and \hb based mass (d), all as a function of the \civ blueshift. The black line is our best-fit linear model, and the shaded region shows the 2-$\sigma$ uncertainties on the slope and intercept. The \ha FWHM have been scaled to match the \hb FWHM using Eq.~\ref{eq:ha2hb}.}  
    \label{fig:correction}
\end{figure*}

In this section we directly compare the \civ and \hans/\hb line widths as a function of the \civ blueshift. 
Because virial BH mass estimates are generally based on the \hb FWHM, we first convert our \ha FWHM measurements to equivalent \hb FWHM using Eq.~\ref{eq:ha2hb}.  
In Fig.~\ref{fig:correction}a and b we show the \civ FWHM relative to both the (\hbns-scaled) \ha FWHM and the \hb FWHM, as a function of the \civ blueshift. 

Employing the same Bayesian fitting framework described in Section~\ref{sec:hahbcomparison}, we fit independent linear models to the \civ FWHM relative to the \ha and \hb FWHM as a function of the \civ blueshift. 
As before, our model has an additional parameter representing any intrinsic scatter in the relationship between the variables which is independent of measurement errors.  
We also tested a model where some fraction of the data points (which is free to vary) are drawn from an outlier distribution, represented by a broad Gaussian centered on the mean of the data. 
We found, however, that the inferred outlier fraction was very low (0.004, corresponding to $\sim$0.7 data points) and so did not include such a component in our model. 

\begin{figure}
    \includegraphics[width=\columnwidth]{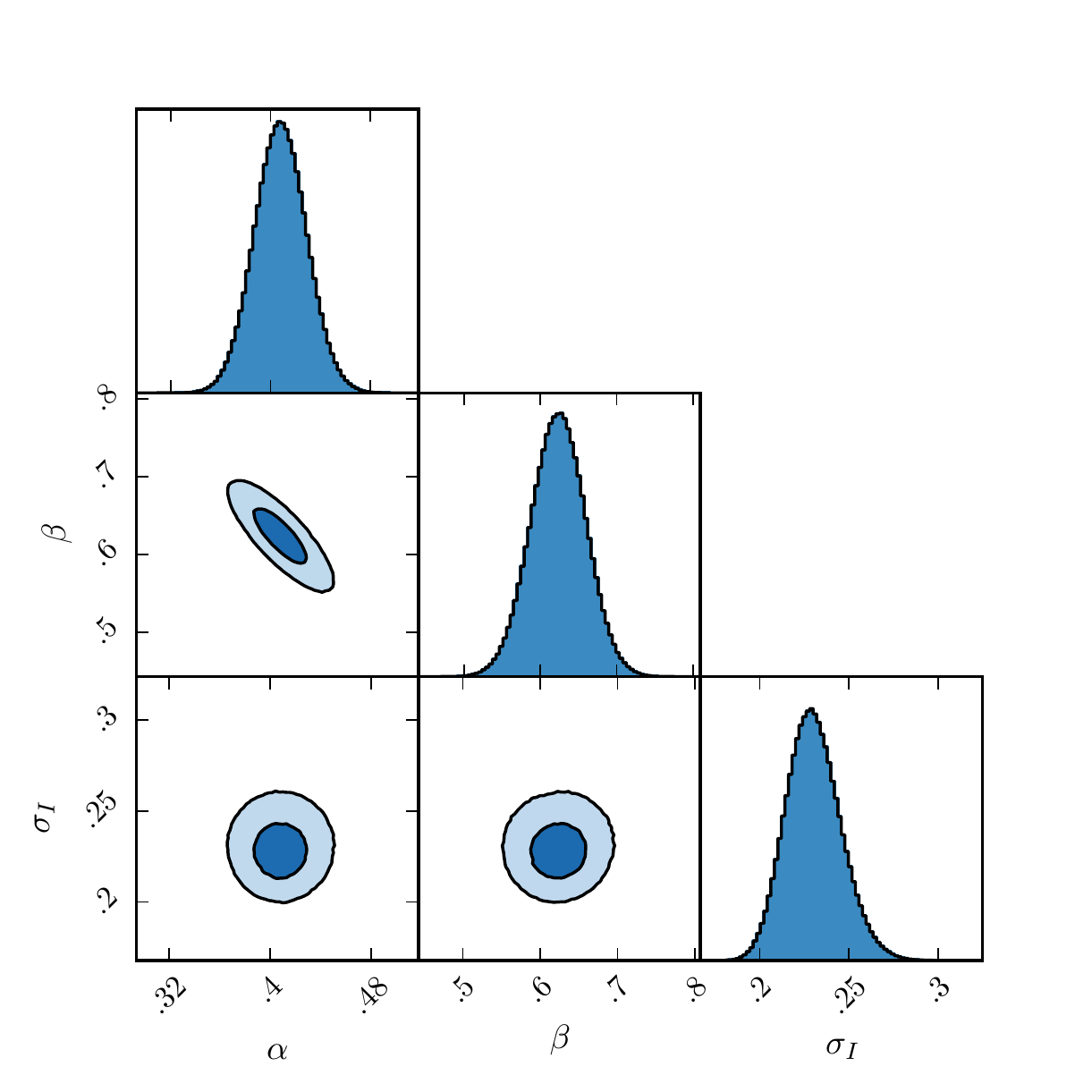} 
    \caption{One- and two-dimensional projections of the MCMC sample of the posterior distribution for a linear fit to the FWHM \civns/\ha ratio as a function of the \civ blueshift. In the two-dimensional projections we show 1- and 2-$\sigma$ contours. The posterior distribution for the linear fit to the FWHM \civns/\hb ratio, which we do not show, has a very similar appearance.} 
    \label{fig:mcmc_parameters}
\end{figure}

In Fig.~\ref{fig:mcmc_parameters} we show the one- and two-dimensional projections of the posterior distribution from the linear fit to the FWHM \civns/\ha ratio. 
The projections from the FWHM \civns/\hb fit (not shown) have very similar appearances.
In Fig.~\ref{fig:correction}a we plot the maximum likelihood model and the 2$\sigma$ uncertainties on the model parameters. 
The maximum likelihood line is given by  

\begin{equation}
    \label{eq:hafwhm}
    \rm{FWHM}(\rm{C}\,\textsc{iv}, \rm{Corr.}) = \frac{\rm{FWHM}(\rm{C}\,\textsc{iv}, \rm{Meas.})}{ (0.41\pm0.02) \left(\frac{\rm{C}\,\textsc{iv}\, Blueshift}{10^3 {\rm \,km\,s^{-1}}} \right) + (0.62\pm0.04)}
\end{equation}

\noindent for the \civns/\ha fit and 

\begin{equation}
    \label{eq:hbfwhm}
    \rm{FWHM}(\rm{C}\,\textsc{iv}, \rm{Corr.}) = \frac{\rm{FWHM}(\rm{C}\,\textsc{iv}, \rm{Meas.})}{ (0.36\pm0.03) \left(\frac{\rm{C}\,\textsc{iv}\, Blueshift}{10^3 {\rm \,km\,s^{-1}}} \right) + (0.61\pm0.04)}
\end{equation}

\noindent for the \civns/\hb fit. 
The intercepts of the two relations are consistent, while the difference between the slopes is only marginally inconsistent given the quoted uncertainties. 

\begin{figure}
    \includegraphics[width=\columnwidth]{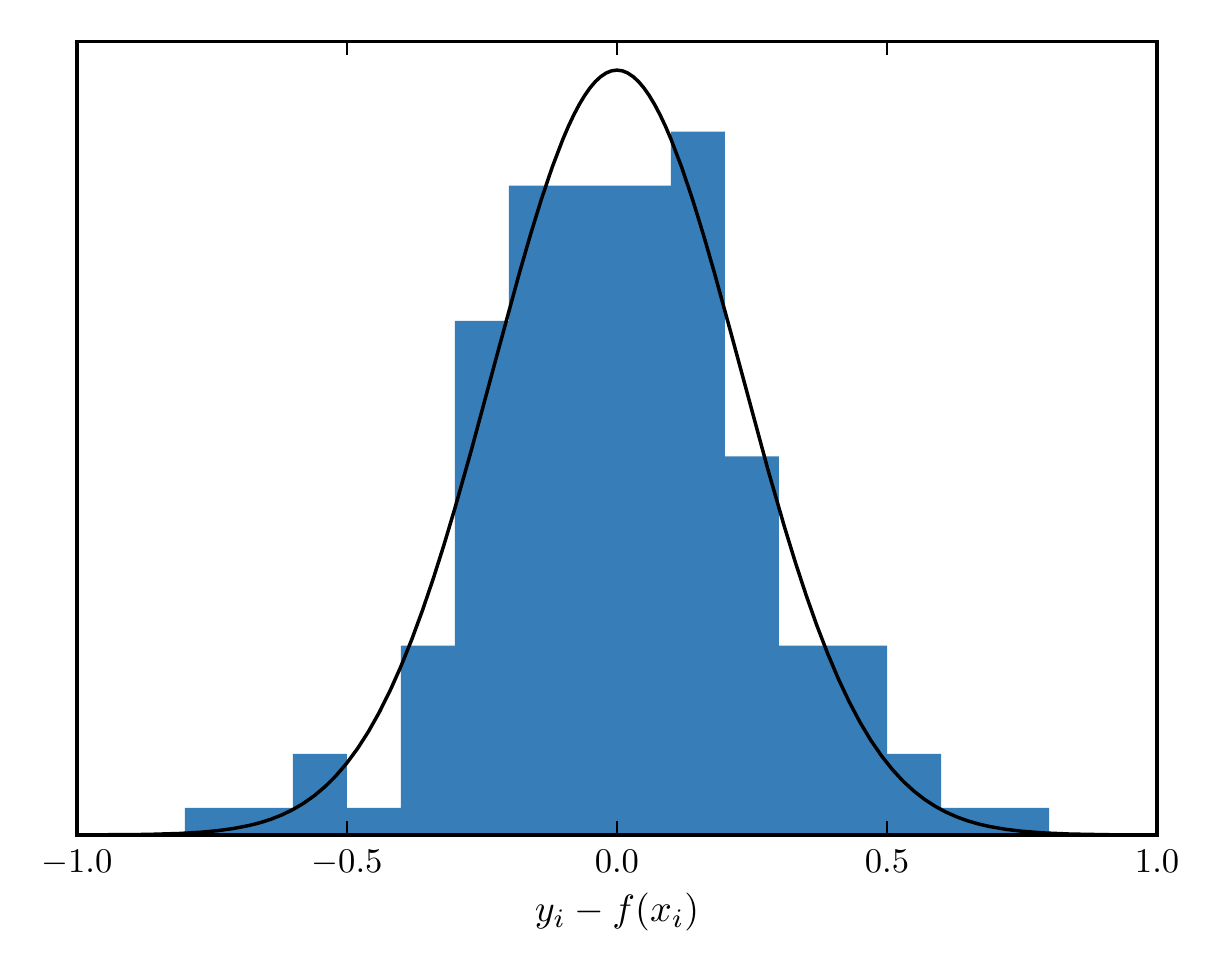} 
    \caption{The distribution of the orthogonal displacement of each data point from the best-fitting linear relationship in the fit to FWHM(\civns)/FWHM(\hans) as a function of the \civ blueshift (blue histogram). The black curve is a Normal distribution with a width equal to the intrinsic scatter in the population inferred from the fit. The two distributions are well-matched, which demonstrates that our model is a good representation of the data and the measurement errors on the data points are small relative to the intrinsic scatter.} 
    \label{fig:intrinsic_scatter}
\end{figure}

The intrinsic scatter in the data about the linear relation we infer is $0.23 \pm 0.02$ and $0.25 \pm 0.02$ for the \ha and \hb fits respectively. 
The intrinsic scatter for the \ha fit is represented by the Normal probability density distribution shown in Fig.~\ref{fig:intrinsic_scatter}. 
In the same figure we show the distribution of the orthogonal displacement of each data point from the best-fitting linear relationship. 
The two distributions are well-matched, which demonstrates that our model is a good representation of the data and the measurement errors on the data points are small relative to the intrinsic scatter.    

The overall (intrinsic and measurement) scatter about the best-fitting model is slightly higher when the \civ line-widths are compared to \hb (0.12 dex) than when compared to \ha (0.10 dex). 
This is likely due, at least in part, to the generally higher S/N of the \ha emission. 
In addition, contributions from the strong \oiii doublet in the vicinity of \hb make de-blending the \hb emission more uncertain. 
As a consequence, for quasars where \ha and \hb are both measured, the mean uncertainty on the \ha FWHM is 130\kms, compared to 340\kms for \hbns. 

In the next section we use both the \ha and \hb lines to calculate unbiased BH masses. 
We use the \ha measurements to derive an empirical \civ blueshift based correction to the \civ masses (Eq.~\ref{eq:masscorrection}) because of the issues related to the accurate modelling of the \hbns-profile just described.  
An extra advantage, which is evident in Fig.~\ref{fig:correction}, is that the \ha sample has a better \civ blueshift coverage. 
However, as can be seen from the similarity of Equations \ref{eq:hafwhm} and \ref{eq:hbfwhm}, our results would not change significantly were we instead to use the \hb sample. 

\subsection{\civ based virial BH mass estimates}

We calculate virial BH mass estimates from \civns, \ha and \hb using the widely-adopted \citet{vestergaard06} scaling relations (their equations 5 and 7 for \hans/\hb and \civ respectively). 
In Figs.~\ref{fig:correction}c and d the \civns-based estimates are compared to the \hans/\hb estimates as a function of the \civ blueshift. 
There is a strong systematic error in the \civns-based masses as a function of blueshift, which is a direct consequence of the FWHM trend described in the previous section. 
The \civ emission-based BH-masses are in error by a factor of more than five at 3000\kms in \civ emission blueshift and the overestimate of the BH-masses reaches a factor of 10 for quasars exhibiting the most extreme blueshifts, $\gtrsim$5000\kms. 

The virial product is the product of the virial velocity squared and the BLR radius \citep[e.g.][]{shen13}, and is proportional to the BH mass. 
We use the corrected \civ FWHM given by Eq.~\ref{eq:hafwhm} as an indicator of the virial velocity, and adopt the same $R-L$ relation for the 1350\,\AA\, continuum luminosity as \citet{vestergaard06} (i.e. $R \propto L^{0.53}$). 
To find the constant scaling factor necessary to transform the virial product in to a BH mass we compute the inverse-variance weighted mean difference between the virial products and the \hans-based masses. 
The virial BH mass can then be expressed in terms of the corrected \civ FWHM and monochromatic continuum luminosity at 1350\,\AA

\begin{equation}
  \label{eq:masscorrection}
  {\rm MBH}(\rm{C}\,\textsc{iv}, \rm{Corr.}) = 10^{6.71}\left( \frac{\rm{FWHM}(\rm{C}\,\textsc{iv}, \rm{Corr.})}{10^3 {\rm \,km\,s^{-1}}} \right)^2 \left( \frac{\lambda {\rm L}_{\lambda} (1350 \text{\AA}) }{10^{44} {\rm \,erg\,s^{-1}}}  \right)^{0.53}
\end{equation}

Given measured \civ emission line FWHM and blueshift, equations 4 and 6 can then be used to provide an unbiased estimate of the quasar BH mass.

\section{Practical application of the \civns-based correction to virial BH-mass estimates}

\subsection{Recipe for unbiased \civ based BH masses}
\label{sec:recipe}

Equations 4 and 6 together provide an un-biased estimate of the virial BH mass given the FWHM and blueshift of \civns, together with the continuum luminosity at 1350\,\AA. 
The FWHM is readily obtained, either directly from the data, or, via the fitting of a parametric model to the \civns-emission line. 
The blueshift -- defined as the bisector of the cumulative line flux -- is also straightforward to measure and our preferred procedure is described in Section~\ref{sec:civ}.
The only potential complication arises in establishing the quasar systemic redshift and hence defining the zero-point for the \civns-blueshift measurement, since both the blueshift and the systemic redshift cannot be determined from \civ alone. 
In practice, when rest-frame optical lines are accessible, as is the case for the quasar sample here, an accurate systemic redshift can be obtained. 
The \oiii doublet and the Balmer lines all have velocity centroids very close to systemic, and the same is true for the broad \mgii doublet. 
For quasars at very high redshifts, $z\sim6$, systemic redshifts can also be derived using the [\ciins] 158 $\mu$m emission in the sub-millimetre band \citep[e.g.][]{venemans16}. 
However, in general, for example in determining the BH-masses of quasars at redshifts $z>2$, if only the rest-frame ultraviolet region is available determining a reliable systemic redshift is non-trivial. 

\citet{shen16b} and our own work shows that there is an intrinsic variation of $\sigma$$\simeq$220\kms in the velocity centroids of the broad-line region relative to a systemic-frame defined by the quasar narrow-line regions. 
As we showed in Paper I, the SDSS DR7 pipeline redshifts are not sufficiently reliable to measure the \civ blueshift accurately because, in part, the \civ emission line itself contributes to the determination of the quasar redshifts (see figure 1 in Paper I).
The redshift-determination scheme of \citet{hewett10} provided much improved redshifts for the SDSS DR7 quasar catalogue, not least because the redshift estimates for the majority of quasars were derived using emission-lines other than the \civns-line itself.
The redshifts for quasars in the SDSS DR10 and DR12 catalogues \citep{paris14,paris16} possess errors of $\simeq$500-750\kms \citep{paris12, font-ribera13}. The impact of low spectrum S/N for fainter quasars in all the SDSS data releases increases the uncertainty further. 
Table~\ref{tab:bhm_error} includes the values for the fractional error in the corrected BH-mass that result from a given error in the determination of the systemic rest-frame. 
For example, the fractional error in the corrected BH mass is 0.39 for a quasar with a 1000\kms\, \civ blueshift when there is a 500\kms\, uncertainty in the quasar systemic redshift.   

Of potentially more significance for studies of BH-masses as a function of quasar and host-galaxy properties are redshift errors that depend on the form of the quasar ultraviolet SED.
The redshifts from \citet{hewett10} still suffer from systematic errors that are correlated with the shape, and particularly the blueshift, of the \civ emission line.
For the \citet{hewett10} redshifts, and ultraviolet emission-line based redshifts in general, quasars with large \civ EW and modest blueshifts have relatively small ($\simeq$300\kms) SED-dependent redshift errors.
Redshift uncertainties as large as $\simeq$1000\kms for such quasars are unusual and the large relative error in the corrected \civ BH-mass given in Table~\ref{tab:bhm_error} is pessimistic. 

Conversely, systematic redshift errors are greatest for quasars with large blueshifts, reaching $\sim$750\kms in the extreme for the \citet{hewett10} values. The associated error in the corrected \civ BH-masses is, however, mitigated somewhat due to the smaller gradient of the MBH(\civns)/MBH(Balmer) relation at large \civ blueshift (see Fig.~\ref{fig:correction}).
A definitive quantification of any systematic SED-dependent errors present in the quasar redshifts contained in the SDSS DR12 catalogue is not yet available but the principal component analysis (PCA) based redshift estimates are expected to be largely free of SED-dependent systematics. 
Given the importance of generating more accurate redshifts for the SDSS DR7 and DR12 quasar samples we will publish a catalogue of more accurate redshifts in due course (see Section~\ref{sec:conclusions}).

\begin{table}
  \centering
  \caption{The fractional error on the corrected BH mass as a function of \civ blueshift for different uncertainties in the quasar systemic redshift.}
  \label{tab:bhm_error}
  \centering
    \begin{tabular}{ccccc} 
    \hline
    \multirow{1}{*}{} & \multicolumn{4}{c}{\civ blueshift (\kms) } \\
    \multicolumn{1}{c}{$\delta$v (\kms)} & 
    \multicolumn{1}{c}{0} &
    \multicolumn{1}{c}{1000} &
    \multicolumn{1}{c}{2000} &
    \multicolumn{1}{c}{4000}  \\
    \hline
    250 & 0.33 &  0.20 &  0.14 & 0.09 \\
    500 & 0.65 & 0.39 & 0.28 & 0.18 \\
    1000 &1.30 & 0.79 & 0.57 & 0.36 \\
    \hline
    \end{tabular}
\end{table}

\subsection{Systematic trends in residuals}

\begin{figure}
    \includegraphics[width=\columnwidth]{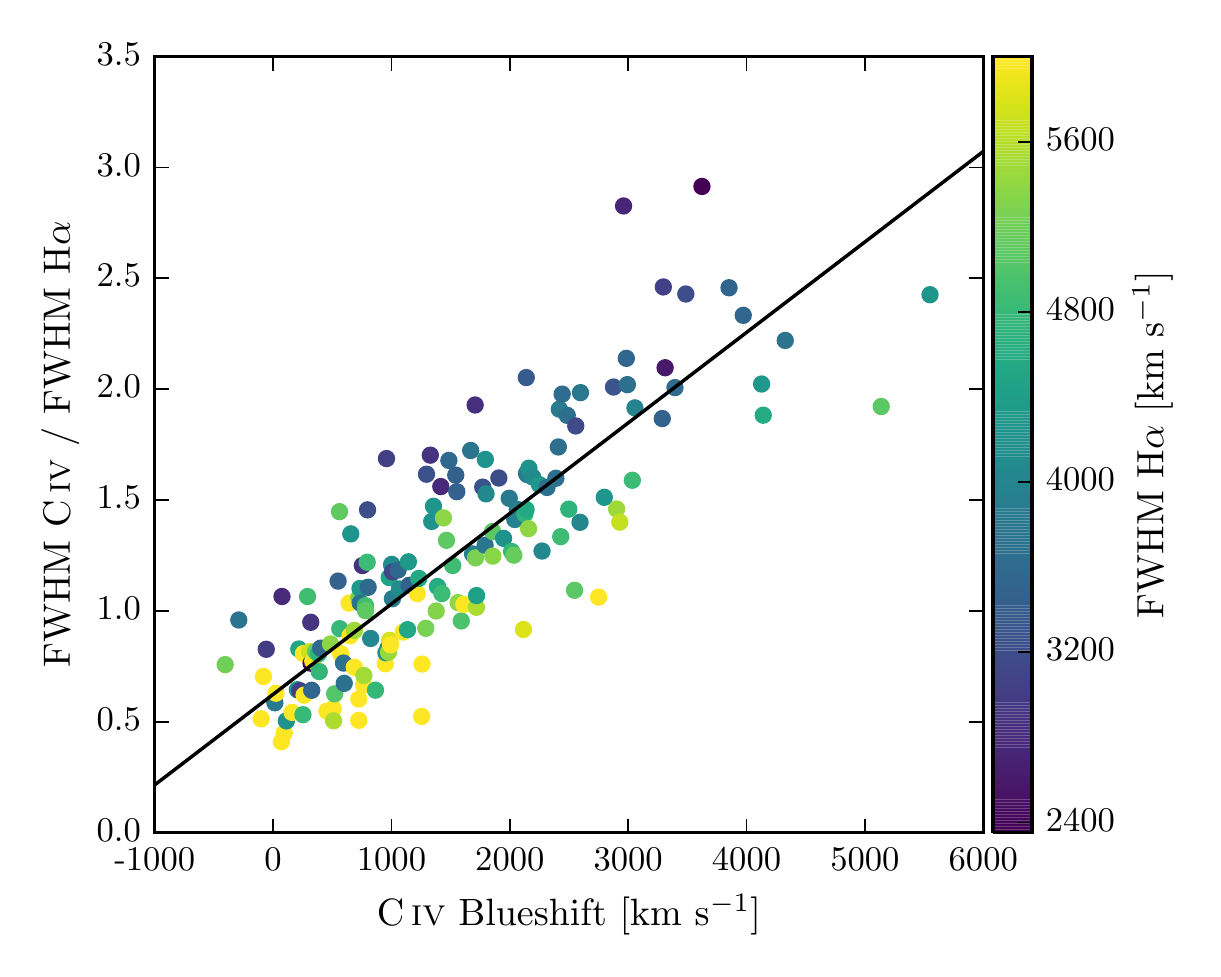}  
    \caption{Same as Fig.~\ref{fig:correction}a, with the marker colour representing the \ha FWHM. At fixed \civ blueshift, there is a clear \ha FWHM dependent systematic in the model residuals.}   
    \label{fig:residuals_ha_fwhm}
\end{figure}

The scatter about the best-fitting line in the \civns/\ha FWHM versus \civns-blueshift relation is $\sim$0.1 dex, an order of magnitude smaller than the size of the \civns-blueshift dependent systematic but, nevertheless, still significant.
With a view to reducing the scatter further, we searched for measurable parameters which correlate with the scatter at fixed \civ blueshift, including the luminosity, redshift, \oiii equivalent width (EW), and \feii EW.
The only significant correlation we find is with the \ha FWHM (Fig.~\ref{fig:residuals_ha_fwhm}).
Quasars with broad \ha lines tend to lie below the relation while quasars with narrow \ha tend to lie above it.
One possibility is that this correlation is simply due to random scatter (either intrinsic or measurement error) in the \ha FWHM which, with the other quasar properties fixed, would naturally produce a correlation between FWHM(\civns)/FWHM(\hans) and FWHM(\hans).
However, the fact that we see no such correlation between the model residuals and the \civ FWHM suggests that the \ha FWHM correlation could be revealing somthing more fundamental. 
The \hans/\hb FWHM is part of `eigenvector 1' (EV1), the first eigenvector in a principal component analysis which originated from the work of \citet{boroson92}.    
While a number of parameters have been considered within the EV1 context \citep[e.g.][]{brotherton99},
Fig.~\ref{fig:residuals_ha_fwhm} suggests that part of the scatter between the Balmer and \civ velocity widths might be attributed to differences in the spectral properties which are correlated with EV1 \citep{marziani13}. 

The residuals and the \ha FWHM also correlate with the shape of the line \citep[FWHM/$\sigma$, where $\sigma$ is the dispersion, derived from the second moment velocity; e.g.][]{kollatschny11,Kollatschny13}. 
The narrow lines are, on average, `peakier' (with FWHM/$\sigma\simeq$ 1) than the broader lines (with FWHM/$\sigma\simeq$ 2).   
The origin of the Balmer-line shape correlation is not clear but one possibility is an orientation-dependence of the \ha FWHM \citep[e.g.][]{shen14}. 
In this scenario quasars with broader emission lines are more likely to be in an edge-on orientation relative to our line of sight.  
    
At radio wavelengths, the morphology of the radio structure, parametrized in terms of `core dominance' is believed, at least in a statistical sense, to be a proxy for the orientation of the accretion disk \citep[e.g.][]{jackson91}.
We matched our sample to the FIRST radio catalogue \citep{white97} in an attempt to identify orientation-dependent signatures.  
Following \citet{shen11}, we classified quasars with matches within 5\,arcseconds as core-dominated, while, if multiple matches were found within 30 arcseconds, quasars were classified as lobe-dominated. 
Twenty core- quasars and six lobe-dominated quasars resulted but no statistically significant differences in the \ha line-widths of the two samples were found. 
It should be noted that the sub-sample of radio-detected quasars is small and the effectiveness of the test is further compromised by the lack of radio-detected quasars at large blueshifts \citep[see figure 14 of][for example]{richards11}.

There are currently very few reverberation-mapping measurements of quasars with large \civ blueshifts.
Looking to the future, the results of the large on-going statistical reverberation mapping projects \citep[e.g.][]{shen15,kingoz15} for luminous quasars at high-redshift will shed new light on the Balmer line emitting region of the BLR for quasars with a range of \civ blueshifts and lead to a greater understanding of the relation between the Balmer line profile and the BH mass. 

\subsection{Effectiveness of the \civ blueshift based correction to BH masses}
\label{sec:effectiveness}

\begin{figure}
    \includegraphics[width=\columnwidth]{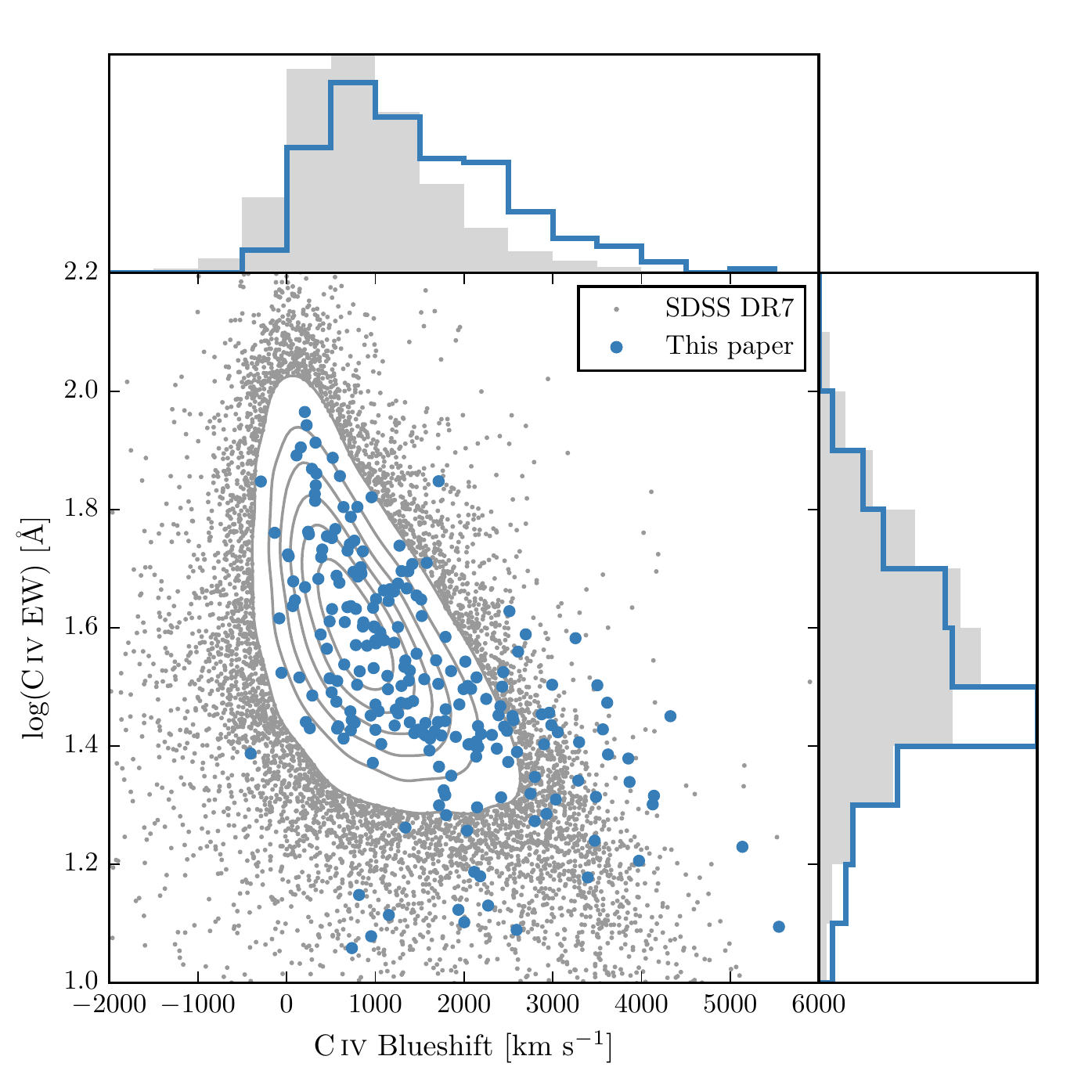} 
    \caption{Rest-frame EW versus blueshift of the broad \civns-emission line for 32,157 SDSS DR7 quasars at $1.6 < z < 3.0$ ({\it grey}) and our sample ({\it blue}). For the SDSS quasars, the systemic redshifts used to calculate the blueshifts are from \citet{hewett10} and \civ emission properties are decribed in Paper I. In regions of high point-density, contours show equally-spaced lines of constant probability density generated using a Gaussian kernel-density estimator. Our sample has very good coverage; the shift to high blueshifts is a result of the high luminosity of our sample in relation to the SDSS sample and the correlation between luminosity and blueshift.} 
    \label{fig:civ_space}
\end{figure}

Figure~\ref{fig:civ_space} demonstrates that our sample has an excellent coverage of the EW-blueshift parameter space in relation to SDSS DR7 quasars at redshifts $1.6 < z < 3.0$. 
The systematic offset to higher \civ blueshifts for our catalogue relative to the SDSS quasars as a whole is a result of the higher mean luminosity relative to the SDSS sample (Fig.~\ref{fig:lzplane}).
Our sample includes 21 quasars with \civ blueshifts $>$3000\kms, and extends to $\sim$5000\kms, i.e. at the very extreme of what is observed in this redshift and luminosity range. 
Our investigation thus demonstrates that the \civns-blueshift based correction derived in this paper is applicable to very high blueshifts. 
Conversely, there are no quasars in our catalogue with \civ blueshifts $\lesssim$0\kms and we caution against extrapolating the correction formula to negative blueshifts. 

\begin{figure}
    \includegraphics[width=\columnwidth]{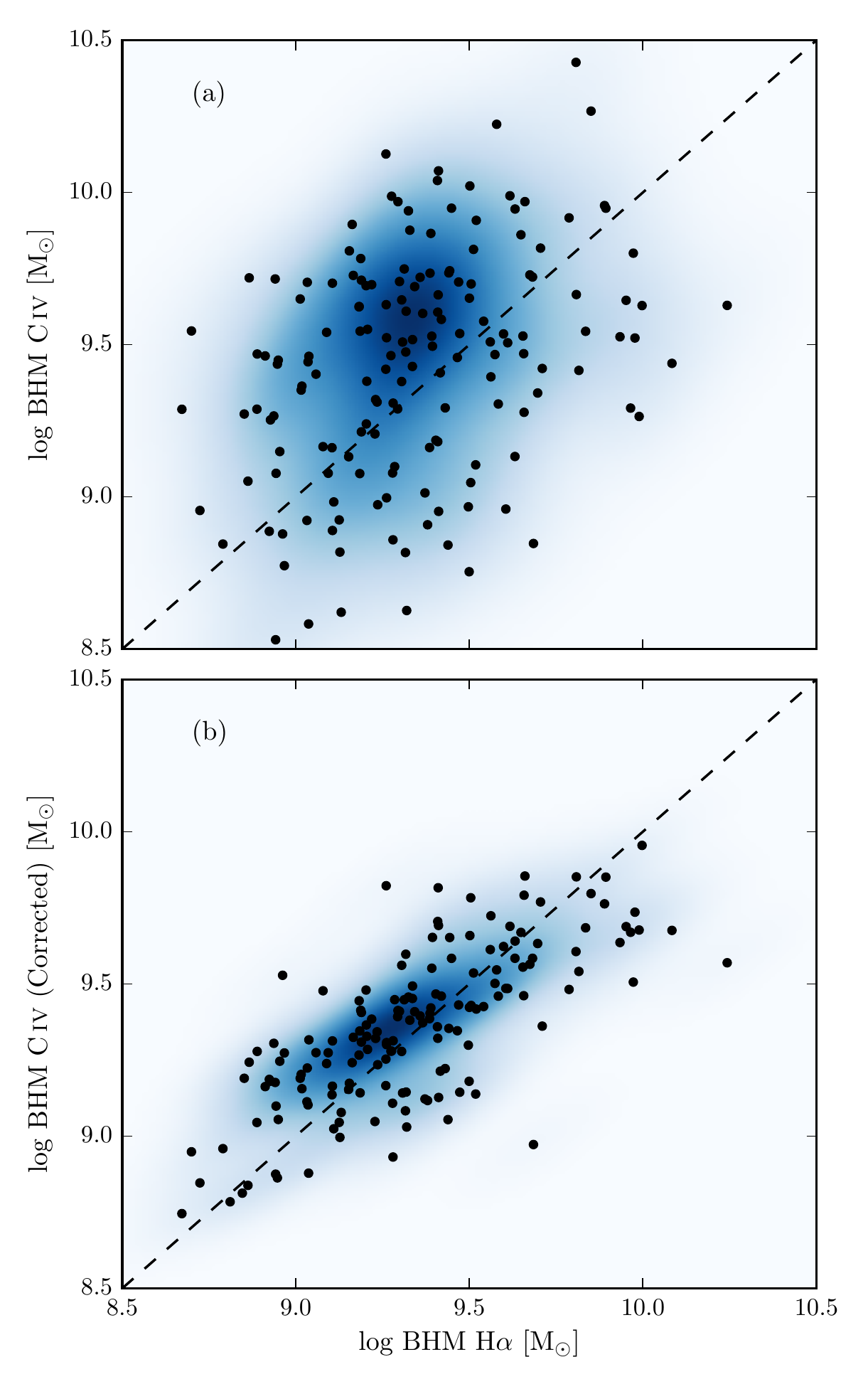} 
    \caption{Comparison of the \civ- and \hans-based BH masses before (a) and after (b) applying the \civ blueshift-based correction to the \civ FWHM. The density of the plotted points (estimated using a Gaussian kernel density estimator) is represented by the colour. The correction to the \civ BH masses decreases the scatter by from 0.4 to 0.2 dex.}   
    \label{fig:bhm_comparison}
\end{figure}

Figure~\ref{fig:bhm_comparison} compares the \civns- and \hans-based BH masses before and after applying the blueshift-based correction to the \civ FWHM.
Before the correction, the correlation between the \civns- and \hans-based BH masses is very weak, and the scatter between the masses is 0.4 dex. 
After correcting the \civ FWHM for the non-virial contribution, the correlation improves dramatically. 
The scatter between the corrected \civns-based masses and the \hans-based masses is reduced to 0.2 dex. 
The scatter is 0.24 dex at low \civ blueshifts ($\sim$0\kms) and 0.10 dex at high blueshifts ($\sim$3000\kms). 

\begin{figure}
    \includegraphics[width=\columnwidth]{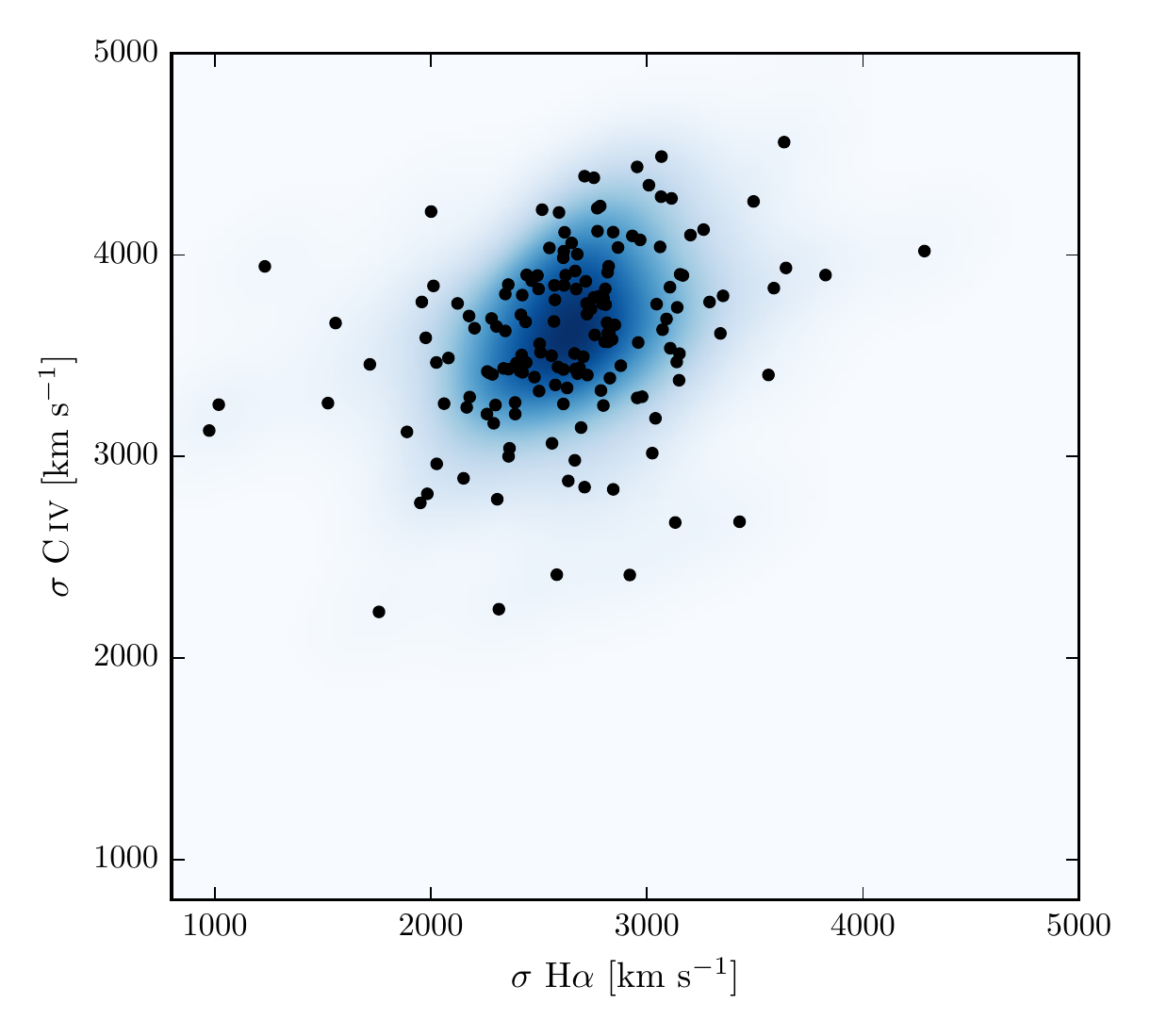} 
    \caption{Comparison of the \civ and \ha line dispersion, $\sigma$. The density of the plotted points (estimated using a Gaussian kernel density estimator) is represented by the colour. Estimating a reliable BH mass from the \civ FWHM and blueshift line is substantially more effective than using the \civ line dispersion with, or without, the line blueshift. The \civ dispersion values are larger than the corresponding \ha measurements by a factor of 1.4 on average, which is consistent with reverberation mapping measurements \citep{vestergaard06}.} 
    \label{fig:dispersion_comparison}
\end{figure}

There has been a considerable amount of attention regarding the relative merits of using the FWHM or dispersion to characterise the velocity width \citep[e.g.][]{denney13}.
As we showed in Paper I, the line dispersion is relatively insensitive to the blueshift and shape of the \civ line. 
Therefore, without the blueshift information, using the line dispersion would yield a more accurate BH mass than the FWHM (Fig.~\ref{fig:dispersion_comparison}). 
The correlation between the \ha and \civ line dispersion is, however, weak. 
The Pearson coefficient for the correlation is 0.36 (and just 0.15 when the \hb measurements are used in place of \hans). 
Furthermore, there is little dynamic range in the line dispersion: the scatter is just 480 and 460\kms for \ha and \civ respectively. The observation suggests that the line dispersion does not fully trace the dynamic range in BH mass present in the quasar population. 
At least part of the reason is that the line dispersion is difficult to measure reliably in current survey-quality data, particularly because of the sensitivity to flux ascribed to the wings of the emission line \citep[e.g.][]{mejia-restrepo16}. 
Figures~\ref{fig:bhm_comparison} and \ref{fig:dispersion_comparison} demonstrate that estimating a reliable BH mass from the \civ FWHM and blueshift line is substantially more effective than using the \civ line dispersion with, or without, the line blueshift. 

\subsection{Comparison to previous prescriptions}

\begin{figure*}
    \includegraphics[width=2\columnwidth]{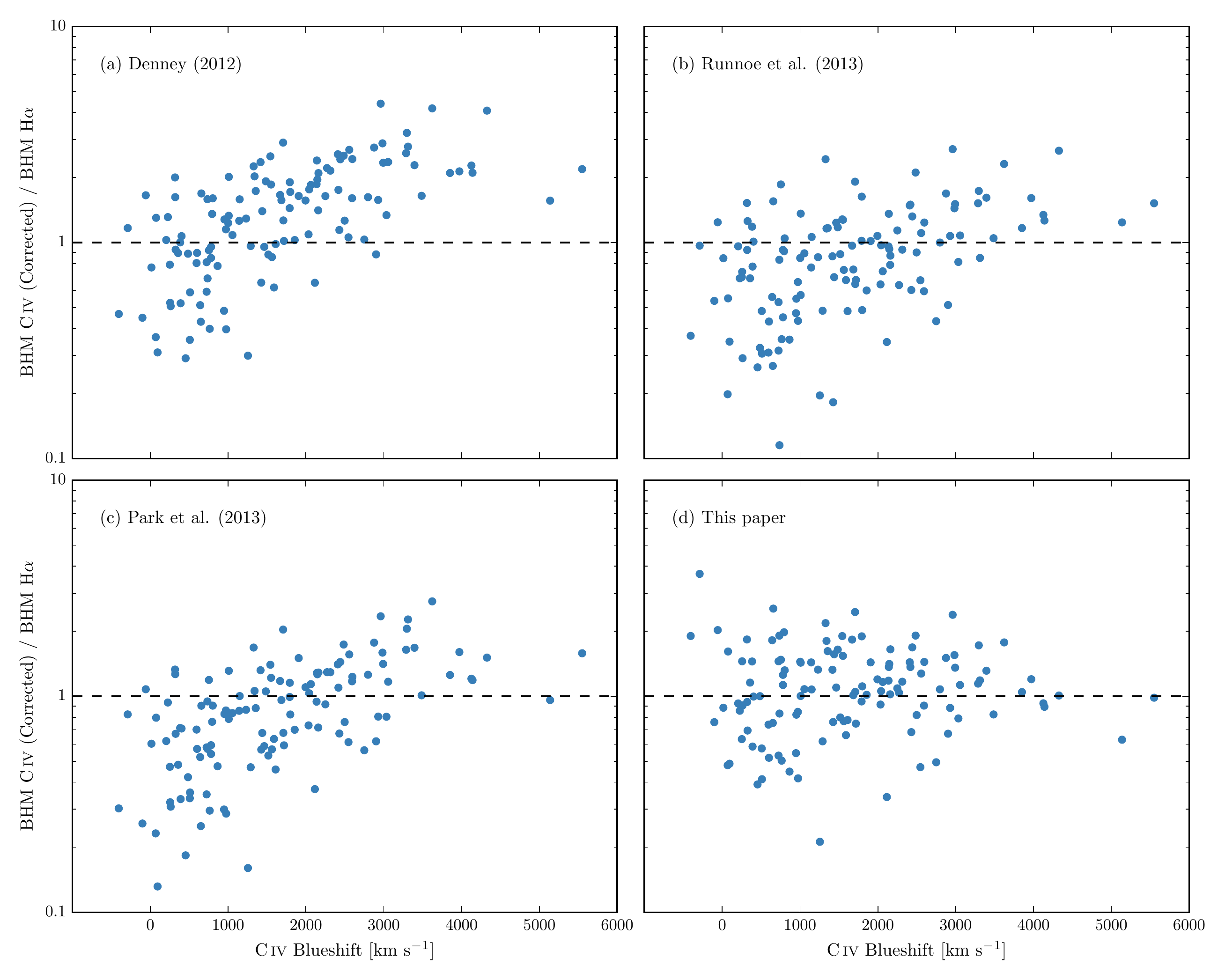}  
    \caption{Comparison of BH mass estimates derived from \civ and \ha as a function of the \civ blueshift. Corrections to the \civns-based masses have been applied based on the shape (FWHM/$\sigma$) of the \civ emission line \citep[a;][]{denney12}, the peak flux ratio of the \siivns+\oiv blend relative to \civ \citep[b;][]{runnoe13}, by significantly reducing the dependence of the derived BH mass on the \civ velocity-width \citep[c;][]{park13}, and based on the \civ blueshift (d; this paper).}
    \label{fig:compare_corrections}
\end{figure*}

In Fig.~\ref{fig:compare_corrections} we compare the \civ blueshift-based correction presented in this paper to various prescriptions which have been proposed in the literature to derive BH masses from the \civ line which are consistent with the masses derived from the Balmer lines. 
In each case we compare the corrected \civns-based masses to the \hans-based masses as a function of the \civ blueshift. 
The correction proposed by \citet{runnoe13} is based on the spectral region at rest-frame wavelengths of $\sim$1400\,\AA \ (see below). 
Therefore, our analysis is based on the 123 quasars which satisfy this requirement. 

In Fig~\ref{fig:compare_corrections}a the \civ BH masses have been corrected using the \civ shape (FWHM/$\sigma$) based correction proposed by \citet{denney12}. 
The correction is not applicable at large \civ blueshifts, since it was calibrated on a sample of low-luminosity AGN which does not include any such objects.
Therefore, while the consistency between the \hans- and \civns-based masses at low \civ blueshifts is improved, at high \civ blueshifts the \civns-based masses remain seriously overestimated.

\citet{runnoe13} used the continuum-subtracted peak flux ratio of the ultraviolet emission-line blend of \siivns+\oiv (at 1400\,\AA) relative to \civns to correct for non-virial contributions to the \civ velocity-width. 
Following \citet{runnoe13}, we measure the peak flux by fitting a model with four Gaussian components (two for each emission line) to the continuum-subtracted flux.
As is evident from Fig.~\ref{fig:civ_space}, a correlation exists between the blueshift and equivalent width of \civns: \civ emission which is strongly blueshifted is typically weak. 
The \siivns+\oiv emission-line blend, however, shows significantly less systematic variation. 
Therefore, the \siivns+\oivns-based correction is quite effective in practice: the systematic bias in the \civ BH masses at large \civ blueshifts is reduced to a factor of $\sim2$ (Fig.~\ref{fig:compare_corrections}b).
However, the \civ based masses are still systematically overestimated at large \civ blueshifts. 

In contrast to the widely-used \citet{vestergaard06} \civns-based virial BH mass calibration, the more recent \citet{park13} calibration significantly reduces the dependence of the derived masses on the emission-line velocity width (from the $V^2$ dependence predicted assuming a virialized BLR to just $V^{0.56}$).
As a consequence, the \civ based masses of the quasars with large \civ blueshifts are much reduced (Fig.~\ref{fig:compare_corrections}c).
However, the systematic error in the \civns-based BH masses as a function of \civ blueshift remains. 

As a comparison, the \civns-based masses shown in Fig~\ref{fig:compare_corrections}d have been corrected using to the \civ blueshift-based procedure presented in this paper. 
No systematic in the BH masses as a function of the \civ blueshift is evident. 

\section{Conclusions}
\label{sec:conclusions}

The main results of this paper are as follows: 

\begin{itemize}
\item We have analysed the spectra of 230 high-luminosity ($10^{45.5}-10^{48}$\,\ergs), redshift $1.5 < z < 4.0$ quasars for which spectra of the Balmer emission lines and the \civ emission line exist.
The large number of quasars in our spectroscopic catalogue and the wide range in \civ blueshifts the quasars possess has allowed us to directly investigate biases in \civns-based BH mass estimates which stem from non-virial contributions to the \civ emission as a function of the \civ blueshift, which, in turn, depends directly on the form of the quasar ultraviolet SEDs \citep{richards11}.
\item The \civ emission-based BH-masses are systematically in error by a factor of more than five at 3000\kms in \civ emission blueshift and the overestimate of the BH-masses reaches a factor of 10 for quasars exhibiting the most extreme blueshifts, $\gtrsim$5000\kms. 
\item We have derived an empirical correction formula for BH-mass estimates based on the \civ emission line FWHM and blueshift.
The correction may be applied using equations 4 and 6 in Section~\ref{sec:correction}.
The large SED-dependent systematic error in \civns-based BH-masses is removed using the correction formulae.
The remaining scatter between the corrected \civns-based masses and the \hans-based masses is 0.24 dex at low \civ blueshifts ($\sim$0\kms) and 0.10 dex at high blueshifts ($\sim$3000\kms). 
This is a significant improvement on the 0.40 dex scatter observed between the un-corrected \civ and \ha BH masses. 
The correction depends only on the \civ line properties - i.e. the FWHM and blueshift - and allows single-epoch virial BH mass estimates to be made from optical spectra, such as those provided by the SDSS, out to redshifts exceeding $z\sim 5$. 
\end{itemize}

As discussed in Section~\ref{sec:recipe}, uncertainties in redshift estimation and hence the definition of the systemic rest-frame for quasars impact on the accuracy of the corrected BH-masses.
Using published redshift estimates, notably those from \citet{hewett10} for the SDSS DR7 quasars and the BOSS PCA-based redshifts from \citet{paris16} for SDSS DR12, the correction formula given in Section~\ref{sec:correction} produces significant improvements to \civns-based BH mass estimates.
In a forthcoming work, Allen \& Hewett (in preparation) will present a new redshift-estimation algorithm that produces redshifts independent of the \civ blueshift and other variations in the ultraviolet SEDs of luminous quasars.
Allen \& Hewett will publish improved redshifts for all quasars in the SDSS DR7 and DR12 which will reduce SED-dependent systematic errors below the apparent inherent dispersion of $\simeq220$\kms associated with broad emission line redshifts \citep{shen16b}.
At the same time we will publish catalogues of unbiased BH masses for both SDSS DR7 and DR12 based on the Allen \& Hewett redshifts. 
The components from the mean-field independent component analysis \citep[see][for an application to astronomical spectra]{allen13} used in the Allen \& Hewett redshift algorithm will also be published.
With these components, if a rest-frame ultraviolet spectrum is available, it will be straightforward to determine the systemic redshift, via a simple optimisation procedure, and hence calculate the \civ blueshift. 

\section*{Acknowledgements}

We thank the anonymous referee for a careful and constructive report that resulted in significant improvements to the paper. 
We are grateful to L. Wisotzki for making optical spectra from the ESO-Hamburg Quasar Survey available for our investigation and Y. Shen for providing us with infrared FIRE and TRIPLESPEC spectra.
We would like to acknowledge the contribution of E. Lusso in collecting the NTT SOFI data. 
We wish to thank R. Williams for help running the Easysinf Sinfoni reduction pipeline and A. Williams for advice on the Bayesian model fitting procedure.  

LC thanks the Science and Technology Facilities Council (STFC) for the award of a studentship. 
PCH acknowledges support from the STFC via a Consolidated Grant to the Institute of Astronomy, Cambridge. 
MB acknowledges support from STFC via an Ernest Rutherford Fellowship. 

Funding for the SDSS and SDSS-II has been provided by the Alfred P. Sloan Foundation, the Participating Institutions, the National Science Foundation, the U.S. Department of Energy, the National Aeronautics and Space Administration, the Japanese Monbukagakusho, the Max Planck Society, and the Higher Education Funding Council for England. The SDSS Web Site is http://www.sdss.org/.

The SDSS is managed by the Astrophysical Research Consortium for the Participating Institutions. The Participating Institutions are the American Museum of Natural History, Astrophysical Institute Potsdam, University of Basel, University of Cambridge, Case Western Reserve University, University of Chicago, Drexel University, Fermilab, the Institute for Advanced Study, the Japan Participation Group, Johns Hopkins University, the Joint Institute for Nuclear Astrophysics, the Kavli Institute for Particle Astrophysics and Cosmology, the Korean Scientist Group, the Chinese Academy of Sciences (LAMOST), Los Alamos National Laboratory, the Max-Planck-Institute for Astronomy (MPIA), the Max-Planck-Institute for Astrophysics (MPA), New Mexico State University, Ohio State University, University of Pittsburgh, University of Portsmouth, Princeton University, the United States Naval Observatory, and the University of Washington.


\bibliographystyle{mn2e}
\bibliography{main} 


\appendix

\section{Comparison to Shen}
\label{appendix:shen}

\begin{figure}
    \includegraphics[width=\columnwidth]{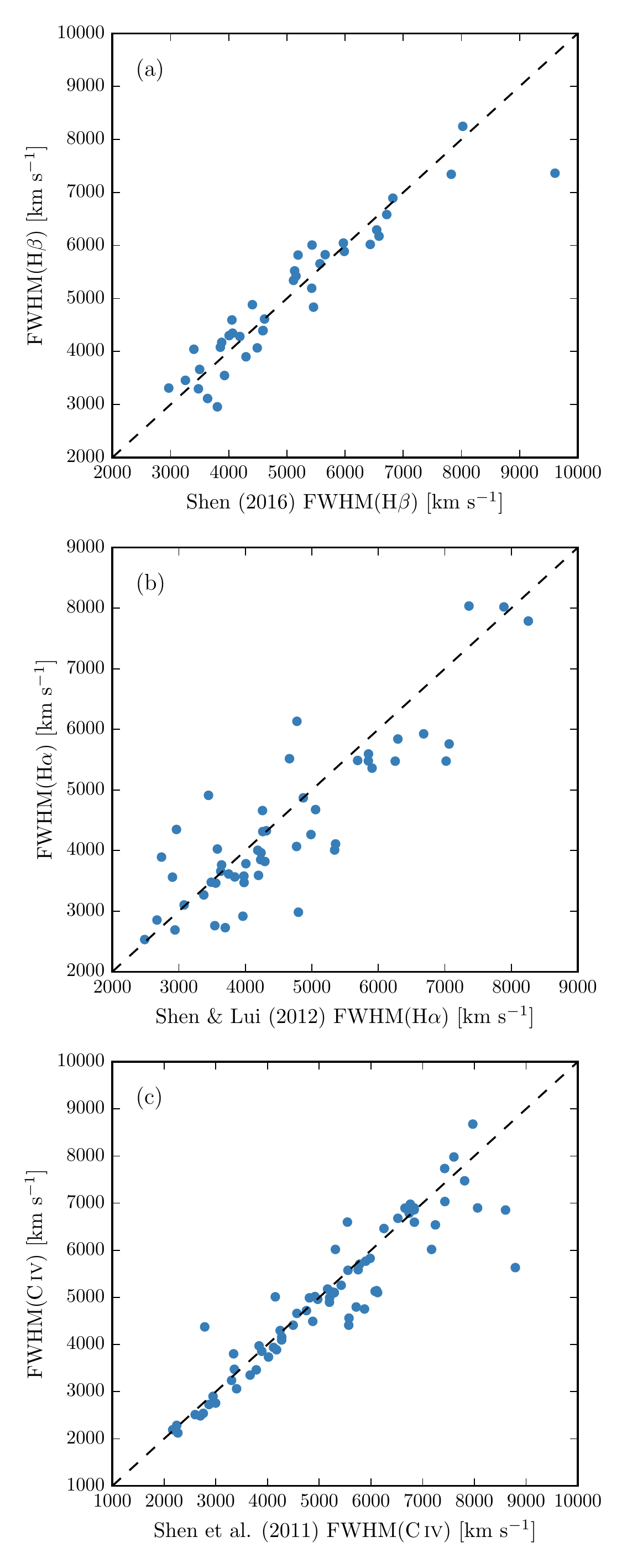} 
    \caption{Demonstration of the effectiveness of our line parameter estimation scheme via a comparison of (a) the \hb FWHM with \citet{shen16a}, (b) the \hb FWHM with \citet{shen12}, and (c) the \civ FWHM with \citet{shen11}.} 
    \label{fig:shen_comparison}
\end{figure}

In this section we verify the accuracy of our parameter estimation scheme by comparing our FWHM measurements to measurements published in \cite{shen11}, \cite{shen12} and \cite{shen16a}. 

The parametric model we fit to the \hbns/\oiii emission region was very similar to the model employed by \citet{shen16a}. 
In Fig.~\ref{fig:shen_comparison}a we plot our \hb FWHM measurements against the measurements published in \citet{shen16a}, for 39 quasars in common to both samples. 
As expected, we observe a very tight correlation, with a scatter of 0.04 dex. 

In Fig.~\ref{fig:shen_comparison}b we plot our \ha FWHM measurements against the measurements published in \citet{shen12}, for 51 quasars in common to both samples.
There is a strong correlation and, although the scatter is larger than for the \hb comparison (0.07 dex), no significant systematic bias. 

Finally, in Fig.~\ref{fig:shen_comparison}c we compare our measurements of the \civ FWHM from the 71 SDSS DR7 spectra in our sample with the measurements published in \citet{shen11}. 
\citet{shen11} fit the \civ line profile with a composite model comprising up to three Gaussian components, whereas we used (up to order six) Gauss-Hermite polynomials. 
Nevertheless, there is a very strong agreement between our measurements, with a scatter of 0.05 dex. 

\section{Spectroscopic catalogue}
\label{appendix:sample}

The observational properties of the sub-samples which make up the quasar catalogue used in this paper are summarised in the following tables. 

\begin{table*}
  \centering
  \caption{Observational properties of the 15 quasars selected from \citet{coatman16} and observed in the near-infrared with the LIRIS spectrograph on the WHT. The columns are as follows: (1) name; (2) date of near-infrared observation; (3) total exposure time (in seconds) of near-infrared observation; (4) S/N measured in the continuum adjacent to \hans, per pixel; (5) source of optical spectra; (6) S/N measured in the continuum adjacent to \civns, per pixel; (7) redshift, taken from \citet{hewett10} unless noted otherwise; (8) photometric magnitude, SDSS $i$-band magnitude unless noted otherwise; (9) radio classification from FIRST: -1 = not in footprint, 0 = radio undetected, 1 = core-dominated radio-detected, 2 = lobe-dominated radio-detected.}
  \label{tab:wht}
  \begin{minipage}{16cm}
  \centering
    \begin{tabular}{ccccccccc}
     \hline
     Name & Date & Exp. (s) & S/N \hans & Opt. Spec. & S/N \civns & $z$ & $m_{i, {\rm SDSS}}$ & Radio \\
     (1) & (2) & (3) & (4) & (5) & (6) & (7) & (8) & (9) \\
     \hline
     J073813.19$+$271038.1 & 2015-04-03 & 1440 & 2.93 & BOSS & 19.47 & 2.4508 & 18.81 & 0 \\
     J074352.61$+$245743.6 & 2015-03-31 & 2520 & 3.41 & SDSS & 5.49  & 2.1659 & 19.07 & 0 \\
     J080651.54$+$245526.3 & 2015-04-04 & 1320 & 1.20 & SDSS & 6.73  & 2.1594 & 18.95 & 0 \\
     J085437.59$+$031734.8 & 2015-04-01 & 720  & 0.95 & BOSS & 17.27 & 2.2504 & 18.42 & 0 \\
     J085856.00$+$015219.4 & 2015-04-01 & 960  & 3.74 & BOSS & 42.33 & 2.1675 & 17.62 & 0 \\
     J110454.73$+$095714.8 & 2015-04-03 & 2880 & 2.27 & BOSS & 12.84 & 2.4238 & 19.12 & 0 \\
     J123611.21$+$112921.6 & 2015-04-01 & 1200 & 3.51 & BOSS & 22.54 & 2.1527 & 18.53 & 0 \\
     J124602.05$+$042658.4 & 2015-04-01 & 1920 & 1.68 & BOSS & 23.52 & 2.4473 & 18.49 & 0 \\
     J133646.87$+$144334.2 & 2015-04-04 & 2160 & 4.97 & BOSS & 9.62  & 2.1422 & 18.84 & 0 \\
     J133916.88$+$151507.6 & 2015-04-01 & 2520 & 0.98 & BOSS & 13.68 & 2.3157 & 18.52 & 0 \\
     J140047.45$+$120504.6 & 2015-04-04 & 1800 & 1.73 & BOSS & 23.56 & 2.1722 & 18.29 & 0 \\
     J153027.37$+$062330.8 & 2015-04-04 & 1680 & 3.83 & BOSS & 19.26 & 2.2198 & 18.62 & 0 \\
     J153848.64$+$023341.0 & 2015-04-02 & 840  & 1.08 & BOSS & 42.09 & 2.2419 & 17.51 & 0 \\
     J161842.44$+$234131.7 & 2015-04-01 & 1200 & 1.93 & SDSS & 15.37 & 2.2824 & 18.50 & 0 \\
     J163456.15$+$301437.8 & 2015-04-04 & 1800 & 2.51 & BOSS & 24.27 & 2.4901 & 18.29 & 0 \\
     \hline
  \end{tabular}
  \end{minipage}
\end{table*}

\begin{table*}
  \centering
  \caption{Observational properties of the quasars taken from \citet{shen12} and \citet{shen16a} and observed in the near-infrared with the TRIPLESPEC spectrograph on the ARC 3.5m telescope (34 quasars) and the FIRE spectrograph on the Magellan-Baade telescope (23 quasars). The typical exposure times were 1.0-1.5 hours for TripleSpec and 45 minutes for FIRE. See Table~\ref{tab:wht} for description of columns.}
  \label{tab:shen}
    \begin{tabular}{cccccccccc}
    \hline
    Name & NIR Spec. & Date & S/N \hans & S/N \hbns & Opt. Spec. & S/N \civns & $z$ & $m_{i, {\rm SDSS}}$ & Radio \\
    \hline
    J002948.04$-$095639.4 & TRIPLESPEC & 10-01-02/10-11-28 & 4.94  & 3.03  & SDSS & 8.72  & 1.6184 & 17.69 & 0 \\
    J004149.64$-$094705.0 & TRIPLESPEC & 10-01-02/10-11-28 & 6.77  &       & SDSS & 20.87 & 1.6287 & 16.97 & 0 \\
    J014705.40$+$133210.0 & TRIPLESPEC & 09-09-09/09-11-07 & 11.11 & 7.95  & BOSS & 24.00 & 1.5950 & 17.09 & -1 \\
    J014944.43$+$150106.7 & TRIPLESPEC & 09-09-09/10-11-28 & 9.94  &       & BOSS & 38.86 & 2.0733 & 17.20 & -1 \\
    J015733.87$-$004824.5 & TRIPLESPEC & 09-11-07/10-11-28 & 2.64  &       & BOSS & 12.68 & 1.5514 & 18.21 & 0 \\
    J020044.50$+$122319.1 & TRIPLESPEC & 10-01-02/10-11-28 & 3.34  &       & SDSS & 11.58 & 1.6538 & 17.83 & 0 \\
    J025905.63$+$001121.9 & FIRE       & 13-12-30          &       & 12.70 & BOSS & 37.49 & 3.3766 & 17.78 & 1 \\
    J030449.85$-$000813.5 & FIRE       & 13-12-29          &       & 18.19 & SDSS & 50.27 & 3.2964 & 17.54 & 0 \\
    J041255.17$-$061210.3 & TRIPLESPEC & 10-01-02/10-11-28 & 8.05  &       & SDSS & 17.86 & 1.6913 & 17.32 & -1 \\
    J074029.82$+$281458.5 & TRIPLESPEC & 09-11-08          & 2.74  &       & BOSS & 16.30 & 1.5452 & 17.45 & 0 \\
    J081344.15$+$152221.5 & TRIPLESPEC & 09-11-08          & 4.94  & 3.90  & BOSS & 12.60 & 1.5453 & 17.54 & 0 \\
    J082146.22$+$571226.0 & TRIPLESPEC & 09-11-08/10-01-04 & 6.90  & 4.64  & SDSS & 28.79 & 1.5458 & 16.86 & 0 \\
    J083850.15$+$261105.4 & TRIPLESPEC & 09-11-08          & 2.50  &       & SDSS & 29.80 & 1.6183 & 16.11 & 0 \\
    J084451.91$+$282607.5 & TRIPLESPEC & 10-12-02          & 1.95  &       & SDSS & 11.25 & 1.5738 & 18.00 & 0 \\
    J091754.44$+$043652.1 & FIRE       & 11-04-27          & 5.64  &       & SDSS & 5.58  & 1.5875 & 18.54 & 0 \\
    J093318.49$+$141340.1 & TRIPLESPEC & 10-01-26          & 5.74  & 3.61  & BOSS & 12.60 & 1.5611 & 17.45 & 0 \\
    J094126.49$+$044328.7 & FIRE       & 11-04-27          & 30.04 & 28.47 & SDSS & 8.91  & 1.5673 & 17.82 & 0 \\
    J094202.04$+$042244.5 & FIRE       & 13-12-28          &       & 19.21 & BOSS & 39.76 & 3.2843 & 17.18 & 0 \\
    J094913.05$+$175155.9 & TRIPLESPEC & 10-01-26          & 11.44 &       & BOSS & 29.48 & 1.6745 & 17.15 & 0 \\
    J100401.27$+$423123.1 & TRIPLESPEC & 10-01-04          & 7.62  &       & BOSS & 33.88 & 1.6656 & 16.77 & 0 \\
    J101447.54$+$521320.2 & TRIPLESPEC & 11-01-24          & 8.18  & 6.60  & BOSS & 22.75 & 1.5524 & 17.33 & 0 \\
    J101504.76$+$123022.2 & TRIPLESPEC & 11-01-24          & 8.12  &       & BOSS & 26.47 & 1.7034 & 17.39 & 0 \\
    J101908.25$+$025431.9 & FIRE       & 13-12-29          &       & 8.18  & SDSS & 26.00 & 3.3791 & 18.12 & 0 \\
    J103456.31$+$035859.4 & FIRE       & 13-12-29          &       & 13.02 & BOSS & 24.39 & 3.3884 & 17.70 & 0 \\
    J104603.22$+$112828.1 & FIRE       & 11-04-26          & 28.03 & 25.55 & SDSS & 8.53  & 1.6067 & 17.79 & 0 \\
    J104910.31$+$143227.1 & TRIPLESPEC & 10-01-26          & 7.52  & 4.91  & BOSS & 14.78 & 1.5395 & 17.74 & 0 \\
    J105951.05$+$090905.7 & TRIPLESPEC & 11-01-24          &       & 9.61  & SDSS & 16.54 & 1.6899 & 16.77 & 0 \\
    J110240.16$+$394730.1 & TRIPLESPEC & 11-02-22          & 8.92  & 5.78  & BOSS & 23.73 & 1.6644 & 17.60 & 0 \\
    J111949.30$+$233249.1 & TRIPLESPEC & 11-01-24          & 10.42 & 7.06  & SDSS & 8.49  & 1.6260 & 17.34 & 0 \\
    J112542.29$+$000101.3 & FIRE       & 11-04-27          & 48.11 & 36.62 & SDSS & 15.55 & 1.6923 & 17.31 & 1 \\
    J114023.40$+$301651.5 & TRIPLESPEC & 10-01-26          & 17.03 & 11.20 & SDSS & 20.97 & 1.5990 & 16.68 & 0 \\
    J122039.45$+$000427.6 & FIRE       & 11-04-27          & 35.43 & 49.55 & SDSS & 27.06 & 2.0484 & 17.19 & 0 \\
    J123355.21$+$031327.6 & FIRE       & 11-04-27          & 25.15 & 23.44 & BOSS & 10.27 & 1.5281 & 17.82 & 0 \\
    J123442.17$+$052126.7 & TRIPLESPEC & 11-05-13          & 8.99  & 6.61  & BOSS & 18.44 & 1.5504 & 17.00 & 0 \\
    J124006.70$+$474003.4 & TRIPLESPEC & 11-02-22          & 8.26  & 5.87  & SDSS & 9.21  & 1.5613 & 17.51 & 1 \\
    J125140.82$+$080718.4 & FIRE       & 11-04-26          & 27.31 & 29.90 & BOSS & 24.26 & 1.6069 & 16.97 & 0 \\
    J135023.68$+$265243.1 & TRIPLESPEC & 11-02-22          & 8.78  & 6.15  & BOSS & 37.07 & 1.6244 & 16.99 & 0 \\
    J135439.70$+$301649.3 & TRIPLESPEC & 11-04-22          & 2.73  &       & SDSS & 10.80 & 1.5529 & 17.68 & 0 \\
    J141949.39$+$060654.0 & FIRE       & 11-04-26          & 40.07 & 37.10 & SDSS & 14.13 & 1.6489 & 17.18 & 0 \\
    J142108.71$+$224117.4 & TRIPLESPEC & 10-05-20/11-05-13 & 20.94 & 18.35 & BOSS & 38.26 & 2.1879 & 16.91 & 1 \\
    J142841.97$+$592552.0 & TRIPLESPEC & 11-04-14/11-04-18 & 5.64  & 3.53  & SDSS & 11.39 & 1.6598 & 17.42 & 0 \\
    J143148.09$+$053558.0 & TRIPLESPEC & 10-05-20          & 14.81 &       & SDSS & 30.41 & 2.0950 & 16.52 & 0 \\
    J143230.57$+$012435.1 & FIRE       & 11-04-27          & 19.05 & 17.39 & SDSS & 12.02 & 1.5422 & 17.64 & 0 \\
    J143645.80$+$633637.8 & TRIPLESPEC & 10-05-20/11-05-13 & 23.75 & 19.66 & SDSS & 35.48 & 2.0662 & 16.75 & 2 \\
    J152111.87$+$470539.1 & TRIPLESPEC & 11-04-22          & 4.06  & 2.60  & BOSS & 12.66 & 1.5169 & 17.53 & 0 \\
    J153859.45$+$053705.3 & FIRE       & 11-04-26          & 39.28 & 29.04 & BOSS & 10.52 & 1.6836 & 17.89 & 0 \\
    J154212.90$+$111226.7 & FIRE       & 11-04-27          & 37.91 & 31.34 & BOSS & 13.80 & 1.5403 & 17.30 & 0 \\
    J155240.40$+$194816.8 & TRIPLESPEC & 11-04-14/11-04-18 & 8.35  & 5.38  & BOSS & 12.92 & 1.6133 & 17.44 & 0 \\
    J160456.14$-$001907.1 & FIRE       & 11-04-26          & 55.98 & 54.20 & SDSS & 12.35 & 1.6364 & 17.08 & 2 \\
    J162103.98$+$002905.8 & FIRE       & 11-07-14          & 2.25  &       & SDSS & 10.35 & 1.6890 & 18.49 & -1 \\
    J171030.20$+$602347.5 & TRIPLESPEC & 11-04-14/11-04-18 & 7.26  & 4.45  & SDSS & 12.02 & 1.5485 & 17.36 & 0 \\
    J204009.62$-$065402.5 & FIRE       & 11-04-26          & 8.05  &       & SDSS & 4.60  & 1.6110 & 18.85 & -1 \\
    J204536.56$-$010147.9 & FIRE       & 11-04-26          & 58.77 & 46.68 & SDSS & 18.66 & 1.6607 & 16.44 & 0 \\
    J204538.96$-$005115.6 & FIRE       & 11-04-27          & 29.69 & 24.34 & SDSS & 6.90  & 1.5896 & 18.11 & 0 \\
    J205554.08$+$004311.5 & FIRE       & 11-04-27          & 21.89 & 21.18 & SDSS & 5.29  & 1.6245 & 18.54 & 0 \\
    J213748.44$+$001220.0 & FIRE       & 11-07-13          & 21.23 &       & BOSS & 19.32 & 1.6699 & 18.07 & 2 \\
    J223246.80$+$134702.0 & TRIPLESPEC & 09-11-07          & 5.97  &       & BOSS & 10.67 & 1.5571 & 17.34 & -1 \\
    \hline
  \end{tabular}
\end{table*}

\begin{table*}
  \centering
  \caption{Observational properties of the quasars taken from the `Quasars Probing Quasars' catalogue. Twenty-two quasars were observed with GNIRS on the Gemini North telescope, four with ISAAC on the VLT, 11 with NIRI also on Gemini North, and nine with XSHOOTER, again, on the VLT. See Table~\ref{tab:wht} for description of columns.}
  \label{tab:qpq}
    \begin{minipage}{16cm}
    \centering 
    \begin{tabular}{ccccccccccc}
    \hline
    SDSS Name & NIR Spec. & Date & Exp. (s) & S/N \hans & S/N \hbns & Opt. Spec. & S/N \civns & $z$ & $m_{i, {\rm SDSS}}$ & Radio \\
    \hline
    J001954.66$-$091316.4 & GNIRS    & 2004-11-26 & 4000  & 10.19 &       & SDSS     & 8.56  & 2.1228                                            & 18.51                                                        & 0 \\
    J011827.98$-$005239.9 & GNIRS    & 2004-11-29 & 2760  & 16.23 & 22.85 & SDSS     & 16.57 & 2.1941                                            & 18.00                                                        & 0 \\
    J013929.51$+$001330.8 & GNIRS    & 2004-12-01 & 4800  & 24.61 & 29.03 & SDSS     & 15.38 & 2.1052                                            & 18.65                                                        & 0 \\
    J014809.64$-$001017.7 & GNIRS    & 2004-11-29 & 2160  & 26.73 & 36.55 & BOSS     & 31.51 & 2.1655                                            & 17.73                                                        & 0 \\
    J020143.48$+$003222.7 & XSHOOTER & 2013-12-03 & 1425 &       & 5.57  & SDSS     & 6.35  & 2.2966                                            & 19.59                                                        & 0 \\
    J020902.86$-$082531.8 & GNIRS    & 2004-10-01 & 4000  & 14.50 & 20.07 & SDSS     & 11.03 & 2.1355                                            & 18.50                                                        & 0 \\
    J032158.40$-$001102.6 & GNIRS    & 2004-11-26 & 4800  & 6.15  &       & SDSS     & 15.86 & 2.1544                                            & 19.31                                                        & 0 \\
    J040954.19$-$041137.0 & GNIRS    & 2004-10-31 & 1800  & 69.99 & 83.20 & SDSS     & 25.51 & 2.1956                                            & 17.68                                                        & -1 \\
    J084159.25$+$392139.9 & NIRI     & 2007-04-24 & 4050  &       & 3.49  & BOSS     & 7.65  & 2.2130\footnote{\label{bosspairs}\citet{paris16}} & 19.52                                                        & 0 \\
    J091208.75$+$005857.3 & GNIRS    & 2004-11-27 & 1800  & 43.87 & 62.40 & BOSS     & 20.52 & 2.1927                                            & 18.25                                                        & 0 \\
    J091432.01$+$010912.4 & XSHOOTER & 2012-01-23 & 1180 & 1.40  & 3.67  & XSHOOTER & 13.00 & 2.1460                                            & 20.53                                                        & 0 \\
    J092747.27$+$290720.6 & NIRI     & 2006-05-15 & 2400  &       & 19.05 & SDSS     & 8.58  & 2.3086                                            & 18.52                                                        & 0 \\
    J093226.34$+$092526.1 & XSHOOTER & 2011-04-04 & 1180 &       & 7.22  & XSHOOTER & 24.27 & 2.4165\footref{bosspairs}                         & 20.20                                                        & 2 \\
    J100246.85$+$002104.0 & GNIRS    & 2005-01-16 & 2400  & 24.18 & 35.56 & SDSS     & 22.65 & 2.1728                                            & 17.84                                                        & 0 \\
    J101859.96$-$005420.2 & GNIRS    & 2005-01-15 & 3045  & 16.87 & 21.99 & BOSS     & 23.04 & 2.1867                                            & 18.37                                                        & 0 \\
    J102906.66$+$020500.0 & GNIRS    & 2005-01-15 & 4000  & 14.68 & 16.16 & SDSS     & 10.82 & 2.1425                                            & 18.58                                                        & 0 \\
    J103325.93$+$012836.3 & GNIRS    & 2005-01-16 & 3720  & 27.30 & 40.76 & BOSS     & 19.58 & 2.1857                                            & 18.43                                                        & 0 \\
    J103857.37$+$502707.9 & NIRI     & 2006-05-09 & 4800  &       & 15.20 & BOSS     & 11.64 & 3.1336                                            & 19.15                                                        & 0 \\
    J104121.89$+$563001.3 & NIRI     & 2007-05-17 & 2100  &       & 20.95 & BOSS     & 33.87 & 2.0535                                            & 18.16                                                        & 0 \\
    J104915.44$-$011038.1 & GNIRS    & 2005-01-15 & 1800  & 33.87 & 43.33 & BOSS     & 34.95 & 2.1249\footref{bosspairs}                         & 17.63                                                        & 0 \\
    J111245.70$+$661215.4 & NIRI     & 2006-06-02 & 6000  &       & 13.64 & BOSS     & 15.41 & 2.2549                                            & 19.14                                                        & -1 \\
    J121427.78$-$030721.1 & GNIRS    & 2005-01-16 & 4000  & 32.52 & 48.77 & SDSS     & 7.09  & 2.1289                                            & 18.46                                                        & 0 \\
    J121558.82$+$571555.5 & NIRI     & 2007-05-23 & 8100  &       & 25.06 & BOSS     & 15.01 & 1.9637                                            & 18.51                                                        & 0 \\
    J122516.79$+$042537.8 & ISAAC    & 2013-06-07 & 3000  &       & 11.59 & BOSS     & 16.64 & 2.4019                                            & 18.87                                                        & 0 \\
    J123143.09$+$002846.2 & XSHOOTER & 2013-04-27 & 2980 &       & 2.51  & XSHOOTER & 6.45  & 3.2051\footref{bosspairs}                         & 19.56                                                        & 0 \\
    J123511.10$-$010829.5 & ISAAC    & 2013-06-05 & 4800  &       & 3.96  & BOSS     & 11.10 & 2.2340\footref{bosspairs}                         & 19.56                                                        & 0 \\
    J123514.36$+$030416.7 & GNIRS    & 2005-01-20 & 4800  & 21.65 & 35.27 & BOSS     & 13.86 & 2.2071                                            & 18.68                                                        & 0 \\
    J134115.56$+$010812.7 & GNIRS    & 2006-02-26 & 1800  & 7.45  &       & BOSS     & 16.56 & 2.2078\footref{bosspairs}                         & 18.92                                                        & 0 \\
    J142758.89$-$012130.4 & GNIRS    & 2006-03-12 & 4200 & 10.73 &       & BOSS     & 11.55 & 2.2781\footref{bosspairs}                         & 19.22                                                        & 1 \\
    J144245.66$-$024250.1 & ISAAC    & 2013-06-05 & 2400  &       & 10.96 & SDSS     & 4.83  & 2.3548                                            & 19.13                                                        & 0 \\
    J151920.75$+$374902.2 & NIRI     & 2007-04-02 & 2700  &       & 24.83 & BOSS     & 23.92 & 2.1119                                            & 17.90                                                        & 0 \\
    J162145.41$+$350807.2 & NIRI     & 2006-04-24 & 3600  &       & 38.54 & BOSS     & 12.49 & 2.0365                                            & 18.57                                                        & 0 \\
    J162548.08$+$264432.5 & NIRI     & 2006-04-17 & 1800  &       & 7.56  & BOSS     & 18.80 & 2.4656\footref{bosspairs}                         & 18.54                                                        & 0 \\
    J162548.80$+$264658.7 & NIRI     & 2006-04-14 & 1800  &       & 24.48 & BOSS     & 54.40 & 2.5313                                            & 17.35                                                        & 1 \\
    J162738.63$+$460538.3 & NIRI     & 2007-05-29 & 9900  &       & 10.47 & BOSS     & 6.44  & 3.8201                                            & 20.13                                                        & 0 \\
    J205954.51$-$001917.4 & GNIRS    & 2004-09-30 & 3000  & 11.26 & 13.92 & SDSS     & 12.18 & 2.1178                                            & 18.49                                                        & 0 \\
    J211832.88$+$004219.0 & GNIRS    & 2004-10-31 & 2400  & 30.88 & 38.71 & SDSS     & 19.85 & 2.1711                                            & 18.00                                                        & 0 \\
    J213438.96$+$000953.7 & ISAAC    & 2013-06-07 & 3600  &       & 5.25  & BOSS     & 16.08 & 2.4131                                            & 19.25                                                        & 0 \\
    J214501.70$-$303122.0 & XSHOOTER & 2013-05-15 & 2980 & 2.57  & 10.87 & XSHOOTER & 21.23 & 2.2190\footnote{\label{2qzpairs}\citet{croom04}}  & 19.59\footnote{\label{bjmag}$B_J$ magnitude (Vega). For a description of the $B_J$ passband see appendix A in \citet{maddox06}.}                 & -1 \\
    J214620.68$-$075250.6 & GNIRS    & 2006-04-09 & 1200  & 16.14 & 21.52 & SDSS     & 7.69  & 2.1204                                            & 19.26                                                        & 0 \\
    J223850.10$-$295612.0 & XSHOOTER & 2013-06-22 & 700  &       & 3.97  & XSHOOTER & 7.09  & 2.4590\footref{2qzpairs}                          & 19.41\footref{bjmag}                                         & -1 \\
    J223850.90$-$295301.0 & XSHOOTER & 2013-05-27 & 700  & 6.62  &       & XSHOOTER & 13.29 & 2.3859\footref{2qzpairs}                          & 19.53\footref{bjmag}                                         & -1 \\
    J225921.59$+$140256.1 & GNIRS    & 2004-10-01 & 2400  & 7.99  &       & SDSS     & 16.96 & 2.1535                                            & 17.97                                                        & -1 \\
    J230959.79$+$005600.9 & XSHOOTER & 2013-07-22 & 2980 & 2.09  &       & XSHOOTER & 3.46  & 2.4133\footref{bosspairs}                         & 19.93                                                        & 0 \\
    J231441.64$-$082406.8 & GNIRS    & 2004-09-30 & 3840  & 23.26 & 31.79 & SDSS     & 11.39 & 2.2044                                            & 18.63                                                        & 0 \\
    J234704.25$+$150146.3 & XSHOOTER & 2012-10-01 & 1180 &       & 5.09  & XSHOOTER & 3.99  & 2.1650\footnote{\citet{prochaska13}}              & 19.89                                                        & -1 \\
    \hline
    \end{tabular}
  \end{minipage}
\end{table*}

\begin{table*}
  \centering
  \caption{Observational properties of the 37 quasars found in the ESO archive and observed in the near-infrared with the SINFONI spectrograph on the VLT. The sample is divided by the ESO programme identification. Optical spectra are from the SDSS/BOSS and the Hamburg/ESO survey (HES). See Table~\ref{tab:wht} for description of columns.}
  \label{tab:sinf}
    \begin{minipage}{16cm}
    \centering 
    \begin{tabular}{cccccccccc}
    \hline
    Name & Date & Exp. (s) & S/N \hans & S/N \hbns & Opt. Spec. & S/N \civns & $z$ & $m_{i, {\rm SDSS}}$ & Radio \\
    \hline
    \multicolumn{10}{c}{083.B-0456(A)} \\
    \hline
    J002018.41$-$233653.8 & 2009-07-07 & 450  & & 17.69 & HES  & 31.16 & 2.2878\footnote{\label{hesinf}\civns-based redshift, measured in Hamburg/ESO spectra} & 14.610\footnote{\label{ksinf} $K$-band magnitude (Vega)} & -1 \\
    J002110.90$-$242247.2 & 2009-07-16 & 600  & & 9.84  & HES  & 31.65 & 2.2534\footref{hesinf}                                                                     & 18.004\footnote{\label{bjsinf} $B_J$-band magnitude (Vega)} & -1 \\
    J002952.12$+$020607.1 & 2009-07-14 & 1200 & & 9.14  & BOSS & 29.33 & 2.3280\footnote{\label{bosssinf}\citet{paris16}}                                           & 15.440\footref{ksinf} & 0 \\
    J004417.12$-$311436.0 & 2009-07-07 & 1800 & & 22.26 & HES  & 23.69 & 2.3438\footref{hesinf}                                                                     & 18.157\footref{bjsinf} & -1 \\
    J005202.51$+$010130.5 & 2009-08-18 & 600  & & 29.08 & BOSS & 40.71 & 2.2775\footref{bosssinf}                                                                   & 14.950\footref{ksinf} & 0 \\
    J005546.77$-$635853.7 & 2009-07-16 & 600  & & 11.68 & HES  & 13.75 & 2.2180\footref{hesinf}                                                                     & 15.280\footref{ksinf} & -1 \\
    J012656.07$-$320810.7 & 2009-08-23 & 450  & & 7.51  & HES  & 37.48 & 2.2127\footref{hesinf}                                                                     & 18.231\footref{bjsinf} & -1 \\
    J024008.24$-$230915.2 & 2009-09-19 & 200  & & 51.58 & HES  & 23.42 & 2.2142\footref{hesinf}                                                                     & 16.571\footref{bjsinf} & -1 \\
    J025644.82$+$001247.1 & 2009-09-24 & 1800 & & 20.58 & BOSS & 39.36 & 2.2636\footref{bosssinf}                                                                   & 15.960\footref{ksinf} & 0 \\
    J034145.86$-$264936.9 & 2009-09-23 & 450  & & 14.63 & HES  & 11.17 & 2.3098\footref{hesinf}                                                                     & 17.193\footref{bjsinf} & -1 \\
    J045754.37$-$181914.7 & 2009-09-20 & 900  & & 23.18 & HES  & 38.4  & 1.5224\footref{hesinf}                                                                     & 17.401\footref{bjsinf} & -1 \\
    J102457.57$-$001740.4 & 2009-04-19 & 450  & & 8.89  & BOSS & 19.16 & 1.4992\footref{bosssinf}                                                                   & 15.300\footref{ksinf} & 0 \\
    J110915.92$-$115449.2 & 2009-07-06 & 900  & & 5.7   & HES  & 9.37  & 2.2937\footref{hesinf}                                                                     & 17.862\footref{bjsinf} & -1 \\
    J120818.84$-$045905.7 & 2009-05-10 & 1800 & & 25.53 & HES  & 11.44 & 2.3005\footref{hesinf}                                                                     & 17.522\footref{bjsinf} & 0 \\
    J121911.34$-$004348.7 & 2009-05-10 & 2400 & & 13.48 & BOSS & 43.34 & 2.2792\footref{bosssinf}                                                                   & 15.350\footref{ksinf} & 1 \\
    J134103.60$-$073947.3 & 2009-07-07 & 450  & & 17.15 & HES  & 15.87 & 2.3450\footref{hesinf}                                                                     & 17.344\footref{bjsinf} & 0 \\
    J144424.38$-$104542.4 & 2009-04-23 & 900  & & 4.88  & HES  & 10.09 & 2.3506\footref{hesinf}                                                                     & 15.260\footref{ksinf} & -1 \\
    J223245.59$-$363202.9 & 2009-06-05 & 600  & & 27.77 & HES  & 37.76 & 2.2636\footref{hesinf}                                                                     & 17.516\footref{bjsinf} & -1 \\
    J225535.43$-$403626.4 & 2009-07-05 & 900  & & 18.39 & HES  & 10.43 & 1.5155\footref{hesinf}                                                                     & 17.989\footref{bjsinf} & -1 \\
    J232539.44$-$065258.6 & 2009-06-26 & 300  & & 25.28 & HES  & 14.77 & 1.4962\footref{hesinf}                                                                     & 16.682\footref{bjsinf} & 0 \\
    \hline
    \multicolumn{10}{c}{090.B-0674(B)} \\
    \hline
    J113334.23$+$130553.2 & 2013-01-05 & 2550                &      & 9.97  & BOSS & 18.97 & 3.6599 & 17.94 & 0 \\
    J115301.60$+$215117.5 & 2013-03-11 & 300($H$)\,2100($K$) & 9.25 & 27.42 & BOSS & 41.43 & 2.3671 & 16.62 & 0 \\
    J130331.28$+$162146.6 & 2013-01-23 & 2100                & 7.97 &       & BOSS & 32.55 & 2.2877 & 18.00 & 1 \\
    J130710.25$+$123021.7 & 2013-02-23 & 2400                &      & 19.08 & BOSS & 28.39 & 3.2087 & 17.56 & 0 \\
    J162014.19$+$103621.1 & 2013-02-24 & 300                 &      & 8.76  & SDSS & 9.80  & 2.0961 & 18.30 & 1 \\
    \hline
    \end{tabular}
  \end{minipage}
\end{table*}

\begin{table*}
  \centering
  \caption{Observational properties of 28 quasars observed in the near-infrared with the SOFI spectrograph on the ESO NTT as part of a programme targeting quasars with archival UVES spectra. See Table~\ref{tab:wht} for description of columns.}
  \label{tab:sofijh}
  \begin{minipage}{16cm}
  \centering
    \begin{tabular}{cccccccccc}
    \hline
    Name & Date & Exp. (s) & S/N \hans & S/N \hbns & Opt. Spec. & S/N \civns & $z$ & $m_{i, {\rm SDSS}}$ & Radio \\
    \hline
    J000344.92$-$232354.8 & 2011-09-18 & 1920 &       & 30.43 & UVES & 65.13  & 2.2800\footnote{\label{zafarsofijh}\citet{zafar13}} & 16.70\footnote{\label{simsofijh}$V$-band magnitude (Vega)}        & -1 \\
    J011143.62$-$350300.4 & 2011-09-19 & 2880 &       & 16.56 & UVES & 76.27  & 2.4060\footref{zafarsofijh}                         & 16.90\footref{simsofijh}                                   & -1 \\
    J012417.38$-$374422.9 & 2011-09-21 & 2880 &       & 19.93 & UVES & 65.1   & 2.1900\footref{zafarsofijh}                         & 17.10\footref{simsofijh}                                   & -1 \\
    J045523.05$-$421617.4 & 2013-03-20 & 4800 & 9.66  & 16.08 & UVES & 83.1   & 2.6610\footref{zafarsofijh}                         & 17.30\footref{simsofijh}                                   & -1 \\
    J082644.70$+$163548.0 & 2013-03-23 & 1920 & 5.68  & 10.58 & SDSS & 27.26  & 2.1942                                              & 17.22                                                      & 0 \\
    J093849.67$+$090509.7 & 2013-03-19 & 3840 & 8.01  & 18.42 & BOSS & 33.85  & 2.2502                                              & 17.33                                                      & 0 \\
    J112442.87$-$170517.5 & 2012-03-05 & 1920 & 18.64 & 35.93 & UVES & 174.27 & 2.4000\footref{zafarsofijh}                         & 16.50\footref{simsofijh}                                   & -1 \\
    J115538.60$+$053050.7 & 2013-03-23 & 3840 &       & 8.1   & BOSS & 22.32  & 3.4758                                              & 17.99                                                      & 0 \\
    J120044.94$-$185944.5 & 2012-03-06 & 2880 & 9.01  & 14.23 & UVES & 67.07  & 2.4530\footref{zafarsofijh}                         & 16.90\footref{simsofijh}                                   & -1 \\
    J120147.91$+$120630.3 & 2012-03-09 & 4800 &       & 11.34 & BOSS & 34.59  & 3.5199                                              & 17.31                                                      & 0 \\
    J121140.59$+$103002.0 & 2013-03-20 & 3120 & 6.56  & 14.42 & BOSS & 29.23  & 2.1977                                              & 17.83                                                      & 0 \\
    J124524.60$-$000938.0 & 2013-03-23 & 2880 & 4.16  &       & BOSS & 30.32  & 2.0902                                              & 17.54                                                      & 0 \\
    J124924.87$-$023339.8 & 2013-03-20 & 4320 & 2.79  &       & BOSS & 33.86  & 2.1208                                              & 17.91                                                      & 0 \\
    J133335.78$+$164903.9 & 2012-03-05 & 2880 & 35.2  &       & BOSS & 48.98  & 2.0840\footref{zafarsofijh}                         & 15.99                                                      & 1 \\
    J134427.07$-$103541.9 & 2012-03-06 & 2880 & 18.96 & 40.74 & UVES & 64.98  & 2.1340\footref{zafarsofijh}                         & 17.10\footref{simsofijh}                                   & -1 \\
    J135038.88$-$251216.8 & 2012-03-05 & 2640 & 7.67  &       & UVES & 80.58  & 2.5780\footref{zafarsofijh}                         & 16.30\footref{simsofijh}                                   & -1 \\
    J140039.00$+$112022.9 & 2012-03-06 & 3840 & 7.46  & 23.24 & BOSS & 41.13  & 2.5790\footnote{\label{bosssofijh}\citet{paris16}}  & 17.15                                                      & 0 \\
    J140445.89$-$013021.8 & 2012-03-09 & 2880 & 4.62  & 3.44  & BOSS & 20.67  & 2.5204                                              & 18.09                                                      & 1 \\
    J143229.25$-$010616.0 & 2012-03-08 & 1920 & 9.5   &       & BOSS & 34.61  & 2.0871                                              & 17.49                                                      & 1 \\
    J143912.04$+$111740.5 & 2012-03-09 & 3840 & 6.55  & 17.12 & BOSS & 21.03  & 2.5918                                              & 18.34                                                      & 1 \\
    J145102.51$-$232931.1 & 2012-03-05 & 1920 & 7.96  & 19.13 & UVES & 57.7   & 2.2150\footref{zafarsofijh}                         & 16.96\footref{simsofijh}                                   & -1 \\
    J155013.60$+$200154.0 & 2013-03-19 & 1920 & 17.31 & 39.74 & SDSS & 25.95  & 2.1956                                              & 16.83                                                      & 1 \\
    J155949.72$+$080517.6 & 2013-03-19 & 2880 & 9.07  &       & BOSS & 37.79  & 2.1770\footref{bosssofijh}                          & 17.28                                                      & 2 \\
    J160222.73$+$084538.4 & 2013-03-22 & 1920 & 7.26  & 16.45 & SDSS & 29.66  & 2.2756                                              & 17.07                                                      & 0 \\
    J161458.30$+$144836.0 & 2012-07-21 & 2640 & 4.51  & 4.27  & SDSS & 28.65  & 2.5664                                              & 17.01                                                      & 0 \\
    J162116.92$-$004250.9 & 2011-09-20 & 3840 &       & 15.6  & SDSS & 29.82  & 3.7285                                              & 17.26                                                      & -1 \\
    J200324.12$-$325145.0 & 2011-09-19 & 6720 &       & 16.49 & UVES & 73.43  & 3.7830\footref{zafarsofijh}                         & 18.40\footref{simsofijh}                                   & -1 \\
    J212912.18$-$153841.0 & 2011-09-22 & 4800 &       & 22.72 & UVES & 54.92  & 3.2680\footref{zafarsofijh}                         & 17.30\footref{simsofijh}                                   & -1 \\
    \hline
    \end{tabular}
  \end{minipage}
\end{table*}

\begin{table*}
  \centering
  \caption{Observational properties of 27 quasars observed in the near-infrared with the SOFI spectrograph on the ESO NTT as part of a programme targeting quasars with large \civ blueshifts. See Table~\ref{tab:wht} for description of columns.}
  \label{tab:sofilc}
  \begin{minipage}{16cm}
    \begin{tabular}{cccccccccc}
    \hline
    Name & Date & Exp. (s) & S/N \hans & S/N \hbns & Opt. Spec. & S/N \civns & $z$ & $m_{i, {\rm SDSS}}$ & Radio \\
    \hline
    J000039.00$-$001803.9 & 2015-09-02 & 3840 & 1.72 &  & SDSS & 8.34 & 2.1342 & 18.81 & 0 \\
    J000500.42$-$003348.2 & 2015-09-01 & 3840 & 3.33 &  & SDSS & 17.17 & 2.1812 & 18.39 & 0 \\
    J000500.53$+$010220.8 & 2015-09-02 & 2880 & 2.99 &  & SDSS & 14.35 & 2.1259 & 18.37 & 0 \\
    J001016.49$+$001227.6 & 2015-09-04 & 2880 & 3.06 &  & BOSS & 28.22 & 2.2766 & 18.38 & 0 \\
    J001919.31$+$010152.2 & 2015-09-04 & 2880 & 1.92 &  & BOSS & 21.79 & 2.3177 & 18.64 & 0 \\
    J002329.60$-$003219.8 & 2015-09-04 & 3840 & 2.06 &  & BOSS & 13.59 & 2.3966 & 18.92 & 0 \\
    J004613.48$+$002358.0 & 2015-09-03 & 3840 & 1.44 &  & BOSS & 15.56 & 2.1250 & 18.69 & 0 \\
    J005202.40$+$010129.2 & 2015-09-02 & 1920 & 7.53 &  & BOSS & 40.71 & 2.2776 & 17.28 & 0 \\
    J005454.84$-$004244.0 & 2015-09-03 & 1920 & 2.63 &  & BOSS & 32.5 & 2.2283 & 17.91 & 0 \\
    J013014.30$-$000639.2 & 2015-08-31 & 2400 & 2.32 &  & BOSS & 34.88 & 2.3992 & 18.16 & 0 \\
    J020327.29$+$003938.1 & 2015-09-04 & 2880 & 3.39 &  & BOSS & 30.0 & 2.3008 & 18.47 & 0 \\
    J020505.80$+$011415.8 & 2015-09-01 & 3840 & 1.21 &  & BOSS & 17.43 & 2.2377 & 19.13 & 0 \\
    J023359.72$+$004938.5 & 2015-09-03 & 1920 & 6.09 & 16.45 & BOSS & 43.43 & 2.5317 & 17.64 & 0 \\
    J024650.93$-$004457.3 & 2015-09-01 & 3840 & 2.03 &  & BOSS & 20.46 & 2.1913 & 18.98 & 0 \\
    J031404.44$-$003947.3 & 2015-09-02 & 2880 & 3.06 &  & SDSS & 26.47 & 2.1127 & 18.51 & 0 \\
    J154432.33$+$020442.0 & 2015-09-02 & 3840 & 1.21 &  & BOSS & 13.96 & 2.3520 & 19.13 & 0 \\
    J212159.04$+$005224.1 & 2015-09-04 & 2880 & 2.59 &  & BOSS & 27.68 & 2.3770 & 18.08 & 0 \\
    J212747.43$+$004929.5 & 2015-09-01 & 1920 & 1.82 &  & BOSS & 49.97 & 2.2567 & 17.53 & 0 \\
    J213125.52$+$001910.4 & 2015-09-03 & 3840 & 2.44 &  & BOSS & 25.85 & 2.1328 & 18.38 & 0 \\
    J213235.95$-$001350.6 & 2015-09-03 & 3840 & 0.99 &  & BOSS & 22.52 & 2.5119 & 18.91 & 0 \\
    J213623.52$-$003410.9 & 2015-08-30 & 4560 & 2.09 &  & BOSS & 26.89 & 2.2296 & 18.29 & 0 \\
    J215615.18$-$003057.8 & 2015-09-01 & 3840 & 1.61 &  & BOSS & 29.35 & 2.3072 & 18.32 & 0 \\
    J215859.35$+$010147.5 & 2015-08-31 & 3840 & 2.37 &  & BOSS & 32.14 & 2.4426 & 18.39 & 0 \\
    J220013.54$-$001912.1 & 2015-09-02 & 6720 & 1.05 &  & SDSS & 5.48 & 2.1287 & 19.57 & 0 \\
    J224026.21$+$003940.1 & 2015-08-31 & 2880 & 3.44 &  & SDSS & 9.2 & 2.1192 & 18.18 & 0 \\
    J225931.72$+$004751.7 & 2015-09-01 & 3840 & 2.59 &  & SDSS & 9.28 & 2.1734 & 18.87 & 0 \\
    J231043.29$-$003151.0 & 2015-09-04 & 2880 & 2.87 &  & SDSS & 6.68 & 2.1715 & 18.66 & 0 \\
    \hline
    \end{tabular}
  \end{minipage}
\end{table*}

\begin{table*}
  \centering
  \caption{Observational properties of 32 quasars observed in the near-infrared with the TRIPLESPEC spectrograph on the Palomar 200-inch Hale telescope. See Table~\ref{tab:wht} for description of columns.}
  \label{tab:triple}
  \begin{minipage}{16cm}
  \centering
    \begin{tabular}{cccllclccc}
    \hline
    Name & Date & S/N \hans & S/N \hbns & Opt. Spec. & S/N \civns & $z$ & Mag. & Radio \\
    \hline
    J005717.37$-$000113.2 & 2011-09-09 &  1.13  &       & SDSS & 8.03  & 2.1555                                             & 19.0  & 0 \\
    J080150.95$+$113455.6 & 2011-04-16 &        & 9.85  & BOSS & 28.66 & 2.3791\footnote{\label{dr12triple}\citet{paris16}} & 18.55 & 0 \\
    J081701.91$+$385936.3 & 2012-04-03 &  3.60  &       & BOSS & 19.87 & 2.2463                                             & 18.66 & 0 \\
    J092914.49$+$282529.2 & 2012-04-04 &        & 7.92  & BOSS & 43.37 & 3.4069                                             & 17.45 & 0 \\
    J092952.17$+$355449.8 & 2011-04-18 &  6.64  & 14.42 & BOSS & 42.43 & 2.1506                                             & 16.85 & 0 \\
    J093337.28$+$284532.4 & 2012-04-04 &        & 7.07  & BOSS & 22.85 & 3.4352\footref{dr12triple}                         & 17.84 & 1 \\
    J094206.96$+$352307.3 & 2011-04-18 &  8.75  & 15.38 & SDSS & 22.48 & 2.0233                                             & 18.04 & 0 \\
    J101001.50$+$403755.5 & 2012-04-03 &  5.64  & 10.56 & BOSS & 20.22 & 2.1906\footref{dr12triple}                         & 18.32 & 1 \\
    J104121.89$+$563001.3 & 2011-04-18 &  6.27  &       & BOSS & 34.04 & 2.0535                                             & 18.16 & 0 \\
    J105158.74$+$401736.7 & 2013-01-29 &  7.6   &       & BOSS & 45.92 & 2.1712                                             & 16.72 & 0 \\
    J110610.74$+$640009.6 & 2011-04-15 &  17.12 & 31.46 & BOSS & 53.52 & 2.2044                                             & 15.97 & 0 \\
    J111350.93$+$401721.4 & 2013-01-29 &  3.32  &       & BOSS & 36.37 & 2.1895                                             & 17.08 & 0 \\
    J112617.40$-$012632.6 & 2011-04-18 &        & 3.57  & BOSS & 15.41 & 3.6253                                             & 18.8  & 0 \\
    J114254.26$+$265457.5 & 2012-04-02 &  8.37  &       & BOSS & 44.79 & 2.6226                                             & 16.62 & 0 \\
    J121117.59$+$042222.3 & 2011-04-15 &  7.58  & 12.08 & BOSS & 34.55 & 2.5487                                             & 17.91 & 0 \\
    J121303.02$+$171423.3 & 2011-04-15 &  7.97  &       & BOSS & 23.67 & 2.5631                                             & 17.63 & 1 \\
    J125353.71$+$681714.3 & 2012-04-04 &        & 3.08  & BOSS & 19.87 & 3.4873                                             & 18.50 & -1 \\
    J130124.73$+$475909.6 & 2012-04-03 &  3.71  &       & BOSS & 21.83 & 2.1843                                             & 18.26 & 0 \\
    J130343.47$+$103113.3 & 2013-01-29 &  3.57  &       & BOSS & 37.28 & 2.1831                                             & 17.28 & 0 \\
    J131011.61$+$460124.5 & 2011-04-16 &  12.6  & 23.37 & SDSS & 33.93 & 2.1425                                             & 16.47 & 0 \\
    J142656.19$+$602550.9 & 2011-04-16 &        & 19.49 & BOSS & 53.54 & 3.1971                                             & 16.23 & 0 \\
    J145408.96$+$511443.7 & 2012-04-02 &        & 5.26  & BOSS & 36.22 & 3.6479                                             & 17.58 & 0 \\
    J154058.70$+$473827.5 & 2012-04-03 &  2.92  & 3.79  & BOSS & 13.23 & 2.5651                                             & 18.98 & 2 \\
    J155233.88$+$491008.3 & 2012-04-03 &  6.37  &       & SDSS & 16.24 & 2.0492                                             & 17.91 & 0 \\
    J155814.51$+$405337.0 & 2011-09-10 &  4.20  &       & BOSS & 20.90 & 2.6385                                             & 18.70 & 0 \\
    J162548.79$+$264658.7 & 2012-04-02 &  4.25  &       & BOSS & 54.40 & 2.5313                                             & 17.35 & 1 \\
    J163412.78$+$320335.5 & 2011-04-16 &  9.41  & 27.99 & BOSS & 37.83 & 2.3531                                             & 17.26 & 1 \\
    J165914.54$+$380900.7 & 2011-09-09 &  2.01  & 10.77 & BOSS & 31.63 & 2.3482                                             & 18.12 & 0 \\
    J172252.98$+$245834.7 & 2012-04-04 &  4.07  & 12.3  & BOSS & 39.83 & 2.2600\footref{dr12triple}                         & 18.06 & -1 \\
    J173352.24$+$540030.5 & 2011-04-16 &        & 5.97  & SDSS & 71.18 & 3.4354                                             & 17.08 & 1 \\
    J223358.87$-$010000.4 & 2011-09-08 &  0.87  &       & BOSS & 24.70 & 2.3286                                             & 18.75 & 0 \\
    J235808.55$+$012507.3 & 2011-09-10 &        & 4.32  & BOSS & 31.38 & 3.3995\footref{dr12triple}                         & 17.38 & 0 \\
    \hline
    \end{tabular}
  \end{minipage}
\end{table*}


\bsp	
\label{lastpage}
\end{document}